\pdfoutput=1
\documentclass[10pt]{iopart}

\usepackage{CJK}
\usepackage{listings}
\usepackage{appendix}
\usepackage{iopams}
\usepackage{graphicx}
\usepackage{color}
\usepackage{scrtime}
\usepackage{bm}
\usepackage{subfigure}
\usepackage[numbers,sort&compress]{natbib} %
\usepackage{hyperref}					
\hypersetup{colorlinks			
	, linkcolor=blue
	, anchorcolor=blue
	, citecolor=blue
	, urlcolor=blue}
\expandafter\let\csname equation*\endcsname=\relax
\expandafter\let\csname endequation*\endcsname=\relax
\usepackage{amsmath,amssymb,amsthm}

\newcommand{\bra}[1]{\langle #1 |}
\newcommand{\ket}[1]{ | #1 \rangle}

\newcommand{\nket}[1]{ | #1 \rangle}
\newcommand{\abs}[1]{|#1|}
\newcommand{\braket}[2]{\langle#1|#2\rangle}
\graphicspath{ {./Figures/}}
\begin{document}

\review{Generalized bulk-boundary correspondence in periodically driven non-Hermitian systems}

\author{Xiang Ji}
\address{Department of Physics, Jiangsu University, Zhenjiang, 212013, China}

\author{Xiaosen Yang} \ead{yangxs@ujs.edu.cn}
\address{Department of Physics, Jiangsu University, Zhenjiang, 212013, China}

\date{\today}

\begin{abstract}
We present a pedagogical review of the periodically driven non-Hermitian systems, particularly on the rich interplay between the non-Hermitian skin effect and the topology. We start by reviewing the non-Bloch band theory of the static non-Hermitian systems and discuss the establishment of its generalized bulk-boundary correspondence. Ultimately, we focus on the non-Bloch band theory of two typical periodically driven non-Hermitian systems: harmonically driven non-Hermitian system and periodically quenched non-Hermitian system. The non-Bloch topological invariants were defined on the generalized Brillouin zone and the real space wave functions to characterize the Floquet non-Hermtian topological phases. Then, the generalized bulk-boundary correspondence was established for the two typical periodically driven non-Hermitian systems.  Additionally, we review novel phenomena in the higher-dimensional periodically driven non-Hermitian systems, including Floquet non-Hermitian higher-order topological phases and Floquet hybrid skin-topological modes. The experimental realizations and recent advances have also been surveyed.  Finally, we end with a summarization and hope this pedagogical review can motivate further research on Floquet non-Hermtian topological physics.
\vspace{2pc}

\end{abstract}
\maketitle
\ioptwocol

\section{Introduction}
In recent decades, the topological materials, which possess gapped bulk states and robust gapless edge states, have attracted a great deal of theoretical as well as experimental interest~\cite{haldane1988model,   kane2005z,kanePhysRevLett2005,bernevig2006quantum,konig2007quantum,  zhang2009topological,hasan2010colloquium, qiRevModPhys2011,alicea2012new,khanikaev2013photonic, beenakker2013search,  susstrunk2015observation, lohse2016thouless, nakajima2016topological,BansilRevModPhys2016, khanikaev2017two, schindler2018higher}. 
These topological phases are beyond Landau's symmetry-breaking theory and cannot be accurately characterized by local order parameters. The essential principle of the topological phases is the bulk-boundary correspondence (BBC): the emergence of robust edge states is the result of the topologically nontrivial bulk~\cite{ fuliangPhysRevLett2007,moore2007topological,schnyder2008classification,  lu2014topological,chiu2016classification,  aidelsburger2015measuring,cooper2019topological, xie2021higher}.

The Hamiltonian of a closed quantum system possesses the property of Hermiticity, which guarantees the probability conservation and reality of eigenvalues. However, the time evolution of open systems should be generated by the non-Hermitian Hamiltonians rather than the Hermitian ones~\cite{bender2007making,huicaoRevModPhys2015,ashida2020non}.
Many phenomena, from open quantum systems to classical wave events, can be well described by non-Hermitian Hamiltonians~\cite{ benderPhysRevLett1998,  Berry2004, YoungwoonPhysRevLett2010, Diehlnaturephys2011,AloisNature2012, leePhysRevX2014, BaoganPhysRevA2014, WiersigPhysRevLett2014,malzardPhysRevLett2015, zhenbonature2015,HosseinNature2017, XuPhysRevLett2017, MenkePRB2017,ZengPRA2017, McDonaldPRX2018,   ShenPhysRevLett2018b,PhysRevACarlstrom2018, ChenYuPhysRevB2018,  LieuPRB2019,YoshidaPhysRevB2019ER, YoshidaPhysRevB2019SPE, chang2019entanglement,midtgaard2019constraints,OkumaPRL2019,   kouPhysRevB2020,XiaoLeiNaturePhysics2020}. 
Especially, the interplay between the non-Hermiticity and the topology has aroused wide attentions in many fields of physics~\cite{KunstPhysRevLett2018,zhesenPhysRevLett2020, zhaihuiNP2020, SukhachovPRR2020,ZhenQianPhysRevLett2020, yang2020exceptional,  PhysRevBYoshida2018, EdvardssonPhysRevB2019,    YangPRB2019,  HochenPhysRevResearch2020,kouPhysRevB2020b,    
imura2020bloch,  HaoranPhysRevLett2020,  zengPRB2020, 
FruchartPhysRevB2016,okuma2023non,
fu2022degeneracy, ding2022non,banerjee2023non}. 
The eigenstates of a non-Hermitian system under the OBC are localized at the boundaries, known as the non-Hermitian skin effect (NHSE)~\cite{zhongwangPhysRevLett2018a, martinez2018topological,prb2019leeAnatomy,prl2019leeTopological,BorgniaPRL2020,YoshidaAPS2020,yu2021generalized,lin2023topological}. Due to the NHSE, the topological edge states can not be characterized by a topological invariant defined by the bulk Bloch states, which indicates the breakdown of Bloch BBC~\cite{ prl2016Lee,zhongwangPhysRevLett2018b,XiongJPC2018, GongPhysRevX2018,LonghiPRL2019}. 
To explain these novel phenomena and establish a generalized BBC in non-Hermitian systems, the non-Bloch band theory~\cite{zhongwangPhysRevLett2018a, ShenPRL2018, yokomizo2019non, Kawabata_2020,HelbigNaturePhysics2020} was developed, in which the Brillouin zone(BZ) was generalized into the generalized Brillouin zone(GBZ). Based on the GBZ, the non-Bloch topological invariants can be defined to characterize the topological edge states and establish the topological classification of non-Hermitian systems~\cite{KawabataPRX2019,nc2019kawabataUnification}. 

Moreover, periodic driving provides a promising platform for generating highly adjustable topological phenomena, even the creation of entirely novel topological states that do not have static equivalents~\cite{jotzu2014experimental,nsture2013rechtsman,LiPRB2019, LimPhysRevLett2008, PhysRevBoka, Soltannaturephy2011,Eckardt_2015,  DrivenRevModPhys2017,YaoPRB2017, XiaosenScientific2018, prb2018zhouNon,HeNatureCommunications2019,  leboeuf1990phase,jiang2011majorana, bermudez2011synthetic, ho2012quantized,   wang2013observation, kundu2013transport,  miyake2013realizing, aidelsburger2013realization,  hu2015measurement, gao2016probing, mahmood2016selective, maczewsky2017observation, mukherjee2017experimental, wang2018floquet,lee2018floquet, rodriguez2018universal, mukherjee2018state, cheng2019observation}. 
For example, the boundaries of the systems can exhibit robust topological edge modes, although the bands are topologically trivial.
Owing to the periodicity in time, the systems have the robust topological zero modes and the $\pi$ modes, which do not have a static analog~\cite{Lindnernaturephy2011, RudnerPhysRevX2013}.
Therefore, the combination of periodic driving and non-Hermiticity will lead to many interesting phenomena without static or Hermitian counter, which have attracted a lot of research interest recently~\cite{ gong2015stabilizing, huang2016realizing, longhi2017floquet, zhan2017detecting, xiao2017observation, chitsazi2017experimental, koutserimpas2018nonreciprocal, turker2018pt, chen2018characterization,  wang2018photonic, leon2018observation, hockendorf2019non, zhou2019dynamical,zhou2019non, liu2020generalized, zhang2020non, pan2020non, wu2020floquet, he2020floquet, mudute2020non, banerjee2020controlling, zhao2019directed, wang2019observation,  cao2021non, zhou2021dual, mittal2021persistence, zhao2021super,chowdhury2021light, ding2021experimental,  zhou2021floquet,prr2021wuFloquet, wu2021floquet_SOTI, weidemann2022topological, prb2022gaoAnomalous, wu2022non, zhou2022q, zhou2022driving, zhao2022quantization,  zhou2023non, ke2023floquet,li2023loss, sun2023photonic, liu2023anomalous, banerjee2023emergent,chowdhury2022exceptional}.
	
In this article, we review the generalized BBC in periodically driven non-Hermitian systems pedagogically. 
In section~\ref{sec:2}, we take a brief review of the non-Bloch band theory and the generalized BBC in the static non-Hermitian Su-Schrieffer-Heeger(SSH) model~\cite{zhongwangPhysRevLett2018a, zhongwangPhysRevLett2018b, yu2021generalized, wang2022amoeba}. 
Based on the GBZ, the non-Bloch topological invariants can be defined to characterize the topological edge states in non-Hermitian systems with the NHSE.
We also introduce the non-Hermitian topological invariants in real space, a powerful tool for characterizing the topological edge states of some systems that are challenging to determine the GBZ.
In section~\ref{sec:3.1}-\ref{sec:3.3}, after reviewing the Floquet theorem, we introduce the generalized BBC in two typical periodically driven non-Hermitian SSH models: harmonically driven non-Hermitian SSH model and periodically quenched non-Hermitian SSH model~\cite{cao2021non, zhou2021dual}. The topological
For the harmonically driven non-Hermitian SSH model, the topological zero modes and $\pi$ modes can be characterized by the non-Bloch winding numbers (NBWNs), which are defined by non-Bloch periodized evolution operators (NBPEOs) based on the GBZ~\cite{cao2021non}.
For the periodically quenched non-Hermitian SSH model, the non-Hermitian winding numbers in real space are defined to characterize the topological edge states~\cite{zhou2021dual}.
Therefore, the generalized BBC was established for the two typical periodically driven non-Hermitian systems, since the non-Bloch topological invariants and open-bulk topological invariants were defined to characterize the Floquet topological edge states accurately.
In section~\ref{sec:3.4}, we take a brief review of the novel phenomena of topological states and NHSE in the higher-dimensional periodically driven non-Hermitian systems, especially describing the Floquet skin-topological effect and Floquet non-Hermitian second-order topological insulators.
In section~\ref{sec:3.5}, we take a concise overview of the experimental realizations and advancements of non-Hermitian as well as periodically driven systems, especially on photonics, acoustic and topological electric circuits.
Finally, we summarize the content of the article and make an outlook on some topics worth discussing in section~\ref{sec:4}.

\section{Non-Bloch band theory}\label{sec:2}
\subsection{Non-Hermitian skin effect}\label{sec:2.1}
As a beginning, let's briefly recall the band theory in Hermitian systems. The eigenstates of a system with the space-translation symmetry are Bloch states which take the form of plane waves modulated by Bloch functions $\psi_n(r)=e^{ikr}u_{n,k}(r)$. 
Here $k$ represents the momentum in BZ and $n$ denotes the band index. 
$u(r)=u(r+a)$ is a period function whose period is the lattice constant $a$. 
The eigenvalues of these Bloch states are labeled as $E_n(k)$ which forms the band structure as momentum $k$.
The Hamiltonian under the PBC and the OBC differs by a boundary term $\delta H$ connecting the boundaries at both ends which causes scattering between different eigenstates. 
The elements of scattering matrix $\bra{n,k}\delta H\ket{n',k'} \sim\abs{\delta H}/L$, where $\abs{\delta H}$ is the strength of the scattering matrix and $L$ is the length of the system, will tends to zero in thermodynamic limits.
Since the size of the actual system is often large, the boundary term can be regarded as a perturbation, which leads to the validity of the Bloch theory under the OBC.

However, non-Hermitian systems under the OBC exhibit novel phenomena that do not have Hermitian counterparts, such as the energy collapse and the NHSE. These are beyond the framework of the Bloch band theory, necessitating the establishment of the non-Bloch band theory.
	\begin{figure}
		\centering
		\includegraphics[width=0.48\textwidth]{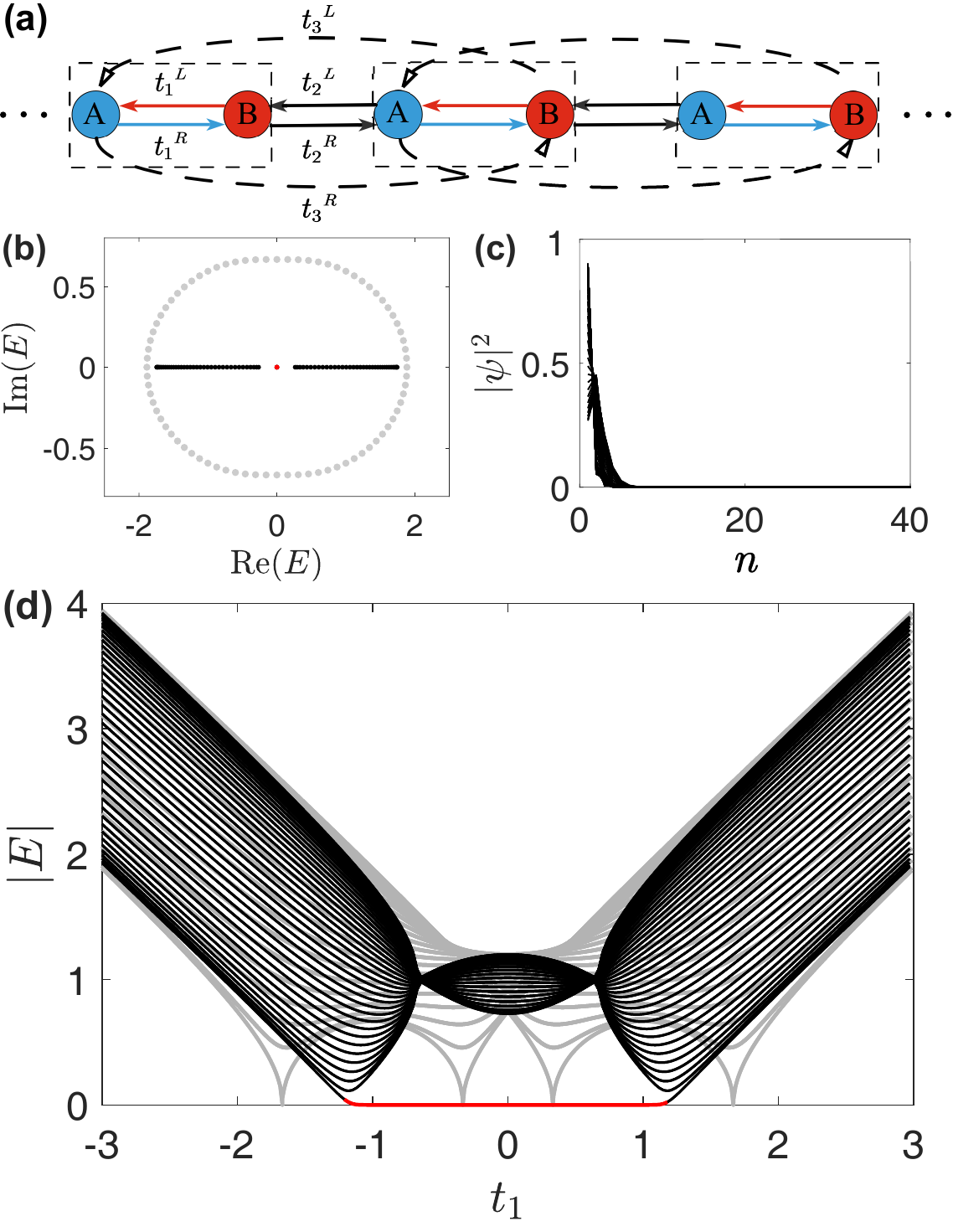}
		\caption{(a) A Sketch of the non-Hermitian SSH model. (b) The energy spectrum of the system under the OBC (black dots) and the PBC (gray dots), respectively. (c) Distribution of eigenstates $\abs{\psi_n}^2=\abs{\psi_{n,A}}^2+\abs{\psi_{n,B}}^2$ under the OBC.
			Here $n$ denotes the index of unit cells. 
			The zero modes are marked in red. (d) The energy spectrum of an open chain with varied $t_1$. Parameters are $L=40$, $t_2=1$,  $\gamma=2/3$, and $t_1=1$ in (a)-(c). Reprinted with permission from~\cite{zhongwangPhysRevLett2018a}, Copyright (Year) by the American Physical Society.}
		\label{fig:NHSSH_1}
	\end{figure}
In particular, a minimum model for illustrating the non-Bloch band theory is the non-Hermitian SSH model~\cite{zhongwangPhysRevLett2018a,ssh1980soliton}. 	A sketch of this model is shown in figure~\ref{fig:NHSSH_1} whose Hamiltonian in real space can be written as
	\begin{align}
		H_{\text{OBC}}=&\sum_{i=1}^{L}(t_1^Lc^\dagger_{i,A}c^{}_{i,B}+t_1^Rc^{\dagger}_{i,B}c^{}_{i,A} )
		\nonumber \\
		& +\sum_{i=1}^{L-1}(t_2^Lc^{\dagger}_{i,B}c^{}_{i+1,A}+t_2^Rc^{\dagger}_{i+1,A}c^{}_{i,B})
		\nonumber \\
		&+\sum_{i=1}^{L-1}(t_3^Lc^\dagger_{i,A}c^{}_{i+1,B} +t_3^Rc^\dagger_{i+1,B}c^{}_{i,A}), \label{eq:NHSSH_obc}
	\end{align}
where $i$ is the position index, $c^{\dagger}$ ($c^{})$ is the creation (annihilation) operator, $t_1^L$, $t_2^L$, $t_3^L$ and $t_1^R$, $t_2^R$, $t_3^R$ are the left and right hopping, respectively. 
For simplicity, only the nearest neighbor hopping is considered, i.e., $t_3^L=t_3^R=0$.
The intracell and intercell hopping amplitudes are taken as $t_1^L=t_1+\gamma$, $t_1^R=t_1-\gamma$ ($t_1>\gamma$) and $t_2^L=t_2^R=t_2$, respectively.
The nonzero  $\gamma$ ($\gamma \neq 0$) breaks the Hermiticity of Hamiltonian.
The Bloch Hamiltonian takes the formula:
	\begin{equation}
		H(k)=n_x\sigma_x+(n_y+i\gamma)\sigma_y,\label{eq:nhssh_hamk1}
	\end{equation}
with $n_x=t_1+(t_2+t_3) \cos k$, $n_y=(t_2-t_3)\sin k$.
The Bloch Hamiltonian has a  chiral symmetry with $\sigma_z H(k) \sigma_z = -H(k)$, which ensures eigenvalues appear in $(E,-E)$ pairs: $E_\pm(k)=\pm\sqrt{n_x^2+(n_y+i\gamma)^2}$. 
	
Figure~\ref{fig:NHSSH_1}(b) depicts the energy spectra of the system under the OBC (black dots) and the PBC (gray dots), respectively. 
The OBC energies are dramatically different from the PBC energies and have collapsed.
As shown in figure~\ref{fig:NHSSH_1}(c), the eigenstates of the system under the OBC exhibit localization at the end of the open chain, which is referred to as the NHSE. 
Notably, the energies of a non-Hermitian Hamiltonian can also have pure real energies~\cite{martinez2018topological}. 
In non-Hermitian systems without NHSE, where gain and loss introduced the non-Hermiticity, the energies can be pure real when the strength of gain and loss is below a threshold~\cite{benderPhysRevLett1998, bender2007making}.
The system possesses parity-time (PT) symmetry~\cite{BaoganPhysRevA2014,GanainyNaturePhysics2018}.
The real-to-complex transition is indicated by the PT symmetry breaking, which is connected to the exceptional points where the eigenvalues and eigenstates coalesce~\cite{feng2014single,miri2019exceptional,ozdemir2019parity}.
As shown in figure~\ref{fig:NHSSH_1}(b) and (c), a non-Hermitian system under OBC can have PT symmetry in the presence of NHSE, whereas under PBC it cannot.
This phenomenon is induced by NHSE~\cite{ol2019LonghiPT,LonghiPhysRevResearch2019} and is referred to as the non-Bloch PT symmetry~\cite{prl2021xueObservation,hu2022geometric}.

Due to the NHSE, the Bloch band theory fails to describe the non-Hermitian system. 
The energies of a non-Hermitian system under the OBC cannot be described by a Bloch Hamiltonian, and the eigenstates are no longer the linear superpositions of Bloch waves.
The Bloch Hamiltonian reveals that the energy gap closes at the exceptional points $(n_x,n_y)=(\pm\gamma,0)$, requiring $t_1= t_2 \pm\gamma$ ($k=\pi$) or $t_1= -t_2 \pm\gamma$ ($k=0$).
Figure~\ref{fig:NHSSH_1}(d) depicts the energy spectra vary with $t_1$ under the PBC (gray lines) and the OBC (black lines), respectively. 
The gap-closing points of energy spectra under the OBC are inconsistent with those under the PBC, indicating the failure of the Bloch BBC.
	
The method of similarity transformation~\cite{zhongwangPhysRevLett2018a} can help us understand the difference between the non-Hermitian SSH model under the OBC and the PBC intuitively. 
The eigenvalue equation of the real-space Hamiltonian is : $H_{\text{OBC}}\ket{\psi}=E\ket{\psi}$ where the eigenvector takes the form of $\ket{\psi}=(\ket{\psi_{1,A}}$,  $\ket{\psi_{1,B}}$, $\ket{\psi_{2,A}}$, $\ket{\psi_{2,B}}$, $\cdots$, $\ket{\psi_{L,A}}$,  $\ket{\psi_{L,B}})$. 
This eigenvalue equation is equivalent to $\bar{H}_{\text{OBC}}\ket{\bar{\psi}}=E\ket{\bar{\psi}}$ by applying  a similarity transformation: 
	\begin{equation}
		\bar{H}_{\text{OBC}}=S^{-1}H_{\text{OBC}}S, \quad \ket{\bar{\psi}}=S^{-1}\ket{\psi},
	\end{equation}
with $S=\text{diag}\{ 1,\ r,\ r,\ r^2,\ \cdots,\ r^{L-1},\ r^{L-1},\  r^{L} \}$ and $r=\sqrt{ \abs{ (t_1-\gamma) / (t_1+\gamma) }}$. Therefore, $\bar H_{\text{OBC}}$ becomes a Hermitian SSH model whose Hamiltonian takes the form:
	\begin{equation}
		\bar H(k)=(\bar t_1+\bar t_2\cos k)\sigma_x+\bar t_2\sin k \sigma_y,
	\end{equation}
where $\bar{t}_{1}=\sqrt{(t_1-\gamma)(t_1+\gamma)}$ and $\bar{t}_{2}=t_{2}$ are the intracell and intercell hopping amplitudes, respectively. 
	
The similarity transformation maps the non-Hermitian Hamiltonian $H_{\text{OBC}}$ into a Hermitian one $\bar{H}_{\text{OBC}}$ without changing the eigenvalues. 
Then, the energy spectrum of $\bar H_{\text{OBC}}$ (or $H_{\text{OBC}}$)  can be captured by $ \bar H(k)$ in thermodynamic limits. The gap-closing points of $\bar H(k)$ are $\bar{t}_{1}=\bar{t}_{2}$, i.e., 
	\begin{equation}
		t_1=\pm \sqrt{t_2^2+\gamma^2},
		\label{eq:t1transition}
	\end{equation}
which are consistent with the transition points of $H_{\text{OBC}}$.
In principle, a topological invariant based on $\bar H(k)$ can be defined to characterize the topological edge states. 
Therefore, the generalized BBC was established in the above non-Hermitian system which exhibits the NHSE.
	
The eigenstates of the non-Hermitian SSH model denoted as $\ket{\psi}=S\ket{\bar{\psi}}$, exhibit localization at the boundaries due to the inclusion of a similarity transformation ($S$), which introduces an exponentially decaying modifier for unit cells ranging from $n=1$ to $n=L$.
Note that the similarity transformation does not apply to the non-Hermitian Hamiltonian under the PBC $H_{\text{PBC}}$. 
Considering the system in equation~(\ref{eq:nhssh_hamk1}), the real-space Hamiltonian is 
$H_{\text{PBC}}=H_{\text{OBC}}+\delta H$ with $\delta H=t_2 c_{L, B}^{\dagger}c_{1, A}+t_2 c_{1, A}^{\dagger} c_{L, B}$.  
Through the similarity transformation, it can be deduced that $S^{-1}H_{\text{PBC}}S=S^{-1}H_{\text{OBC}}S+S^{-1}\delta H S$, in which the boundary term $S^{-1}\delta H S=r^{-L}t_2 c_{L,B}^{\dagger}c_{1,A}^{}+r^{L}t_2 c_{1,A}^{\dagger} c_{L,B}^{}$ increases exponentially with the size of the system. 
Thus, $\delta H$ cannot be regarded as a perturbation, indicating the sensitivity of the non-Hermitian system to the boundaries. This is also the reason why $H_{\text{PBC}}$ and $H_\text{OBC}$ have drastically distinct energy spectra.

\subsection{Generalized Brillouin zone and non-Bloch topological invariants}\label{sec:2.2}
The similarity transformation reveals the significant influence of the boundary conditions on non-Hermitian systems and explains the phenomena of the energy collapse and the NHSE intuitively.
However, it is not general for all one-dimensional (1D) non-Hermitian systems, e.g., it is inapplicable for the model in equation~(\ref{eq:NHSSH_obc}) when $t_1^L\neq t_1^R$, $t_3\neq 0$.
The general theory is the non-Bloch band theory~\cite{zhongwangPhysRevLett2018a}.
For a more intuitive display, the case of $t_3=0$ is taken first in a more generalizable way. Then theory can be applied to $t_3\neq 0$. 
From the eigenvalue equation of the real-space Hamiltonian, it can be derived that
\begin{align}
	t_2\psi_{n-1,B}+(t_1+\gamma)\psi_{n,B}=E\psi_{n,A},
	\nonumber\\
	(t_1-\gamma)\psi_{n,A}+t_2\psi_{n+1,A}=E\psi_{n,B}.
	\label{bulkeigen}
\end{align}
Inspired by the similarity transformation, $\ket{\psi}$ can be proposed with $\ket{\psi}=\sum_j\ket{\phi^{(j)}}$. 
Each $\ket{\phi^{(j)}}$ has the exponential form (temporarily ignoring  $j$ index): $(\phi_{n,A},\phi_{n,B}) = \beta^{n}(\phi_A,\phi_B)$,
which satisfies
\begin{align}
[(t_1+\gamma)+t_2\beta^{-1}]\phi_{B}   &=E\phi_{A}, 
\nonumber\\
[(t_1-\gamma)+t_2\beta]\phi_{A}   &=E\phi_{B}.
\label{eq:nhssh_ansatz}
\end{align}
Then, the following formula can be obtained:
\begin{equation}
\label{eq:nhssh_eigen}
[(t_1-\gamma)+t_2\beta][(t_1+\gamma)+t_2\beta^{-1}]=E^2.
\end{equation}
This equation has two roots:
\begin{equation}
	\beta_{1,2}(E)=\frac{-B \pm \sqrt{B^2-4 t_2^2\left(t_1^2-\gamma^2 \right)}}{2 t_2\left(t_1+\gamma\right)},\label{eq:nhssh_roots}
\end{equation}
with $B=t_1^2+t_2^2-\gamma^2-E^2$.
These two roots satisfy 
\begin{equation}
	\beta_1\beta_2= \frac{t_1-\gamma}{t_1+\gamma}
	\label{eq:nhssh_beta1beta2}
\end{equation}
Restoring the $j$ index, $\phi_A^{(j)}$ and $\phi_B^{(j)} $ satisfy
\begin{align}
		\phi_A^{(j)} & =\frac{E}{t_1-\gamma +t_2 \beta_j} \phi_B^{(j)},
		\nonumber \\
		\phi_B^{(j)} & =\frac{E}{t_1+\gamma +t_2 \beta_j^{-1}} \phi_A^{(j)},
		\label{eq:nhssh_linear}
\end{align}
which are equivalent due to equation~(\ref{eq:nhssh_eigen}). 

The two roots of the characteristic equation imply that the system has two separate exponential wave functions.
The wave functions in real-space can be written as a linear superposition of these two wave functions:
\begin{equation}
	\label{}
	\begin{aligned}
		\begin{pmatrix}
			\psi_{n,A} \\
			\psi_{n,B}\\
		\end{pmatrix}
		=\beta_1^{n}
		\begin{pmatrix}
			\phi_A^{(1)} \\
			\phi_B^{(1)}\\
		\end{pmatrix}
		+\beta_2^{n}
		\begin{pmatrix}
			\phi_A^{(2)} \\
			\phi_B^{(2)}\\
		\end{pmatrix}
	\end{aligned},
\end{equation}
which should satisfy the boundary conditions
\begin{align} \label{boundary}
 (t_1+\gamma)\psi_{1,B}-E\psi_{1,A}=0,
 \nonumber \\
 (t_1-\gamma)\psi_{L,A}-E\psi_{L,B}=0.
\end{align}
In connection with equation~(\ref{eq:nhssh_linear}), the condition for the system to have a non-zero solution is
\begin{align}
\beta_1^{L+1}(t_1-\gamma+t_2\beta_2)=\beta_2^{L+1}(t_1-\gamma+t_2\beta_1). 
\label{eq:nhssh_L} 
\end{align}  
In thermodynamic limit($L\rightarrow \infty$), equation~(\ref{eq:nhssh_L}) leads to $|\beta_1|=|\beta_2|$ for the bulk eigenstates as the solution of $|\beta_1|\neq |\beta_2|$ does not conform to physical facts. 
Combining with equation~(\ref{eq:nhssh_beta1beta2}), we have
\begin{equation}
	\left|\beta_1(E)\right|=\left|\beta_2(E)\right|=r \equiv \sqrt{\left|\frac{t_1-\gamma }{t_1+\gamma }\right|}.\label{eq:nhssh_betaA}
\end{equation}
 
In the similarity transformation, the same $r$ is employed.
All the solutions of $\beta$ satisfied equation~(\ref{eq:nhssh_betaA}) form the GBZ.
According to equation~(\ref{eq:nhssh_betaA}), we take $\beta=re^{ik}$ ($k\in[0,2\pi]$) in equation~(\ref{bulkeigen}) to obtain the energy spectra:
\begin{align}
	E^2(\beta) = & t_1^2+t_2^2 -\gamma^2 +t_2 \sqrt{|t_1^2-\gamma^2|}
	\nonumber \\
	&\cdot[\text{sgn}(t_1+\gamma)e^{ik} +\text{sgn}(t_1-\gamma)e^{-ik}]. \label{eq:nhssh_ebeta} 
\end{align}  
The energies satisfying this equation will coincide with the OBC energy spectra. 
The spectra are real when $|t_1|>|\gamma|$, which indicates the non-Bloch PT symmetry of the system.
The gap-closing points correspond to the topological phase transition and can be determined by equation~(\ref{eq:nhssh_ebeta}).
	
More generally, we consider an arbitrary 1D non-Hermitian tight-binding model~\cite{zhongwangPhysRevLett2018a, zhongwangPhysRevLett2018b, yokomizo2019non}, whose Hamiltonian can be represented as:
	\begin{equation}
		H=\sum_{i,j=1}^{L}\sum_{a,b}^{}t_{i-j}c_{i,a}^{\dagger}c_{j,b}^{},
	\end{equation} 
where $i,\ j$ are the position indices and $a,\ b$ denote the intracell degrees of freedom. 
The hopping amplitude $t_{i-j}$ depends on the spatial distance $i-j$. 
The Hermiticity is broken by the nonreciprocal hopping with $t_{i-j}\neq t_{j-i}$.
The Bloch Hamiltonian can be written as:
	\begin{equation}
		H(k)=\sum_n t_n e^{ink}.
	\end{equation}
	
The non-Bloch Hamiltonian can be obtained by replacing the Bloch factor $e^{ik}$ with the non-Bloch factor $\beta=e^{ik'}=re^{ik}$ ($k\in\mathbb R$) with complex-valued wave vectors $k'=k+i\ln r$. The values of $\beta$ satisfy the characteristic equation:
\begin{equation}
\det[E-H(\beta)]=0,\label{eq:SSH_CE}
\end{equation}
which is a $2M$th-order polynomial with $\det[E-H(\beta)]=a_{-M} \beta^{-M}+\cdots+a_{M}\beta^{M}$. All the solutions can be sorted as $|{\beta_1(E)}| \leq |\beta_{2}(E)| \leq \cdots \leq |\beta_{2M}|$. The values of $\beta$ form the GBZ with the continuous band condition:
\begin{equation}
	|\beta_M| = |\beta_{M+1}|.
\end{equation}
All the $\beta_M$ and $\beta_{M+1}$ form closed curves, which can be denoted as $C_{\beta}$ ~\cite{zhongwangPhysRevLett2018a,yokomizo2019non}. 
  
\begin{figure}
	\centering
	{\includegraphics[width=5.0cm, height=4.1cm]{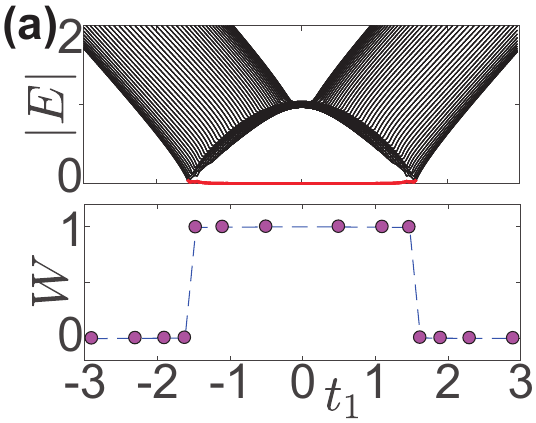}}
	{\includegraphics[width=4.6cm, height=4.4cm]{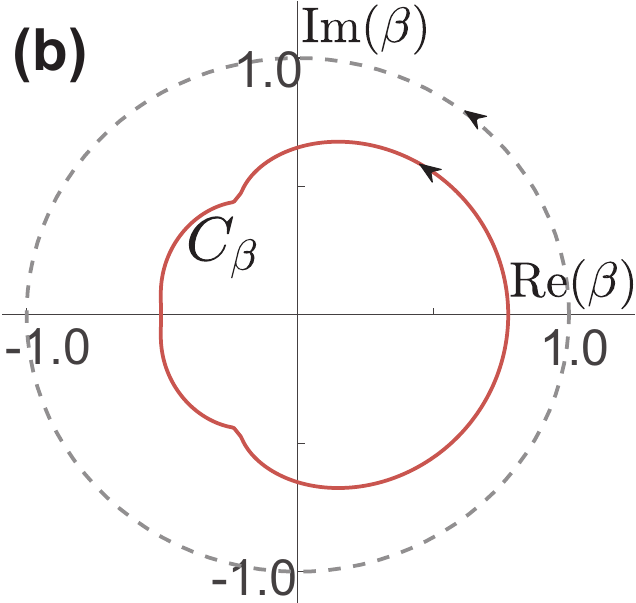}}
	\caption{ The case of $t_3\neq 0$. (a) Energy spectra with varied $t_1$ of an open chain and the corresponding NBWNs. The transition points correspond to $t_1\approx\pm 1.56$. (b) The GBZ ($C_\beta$) of $t_1=1.1$. Parameters are $t_2=1$, $\gamma=2/3$, $t_3=1/5$. Reprinted with permission from ~\cite{zhongwangPhysRevLett2018a}, Copyright (Year) by the American Physical Society. }\label{fig:nhssh_gbz}
\end{figure}  
  	
Based on the GBZ, energy spectra of bulk under the OBC can be described by the non-Bloch Hamiltonian. 
In addition, topological invariants can be defined in terms of GBZ to characterize the topological edge states. The non-Bloch Hamiltonian of equation~(\ref{eq:nhssh_hamk1}) is

\begin{equation} 
H(\beta)= (t_1-\gamma + \beta t_2 )\sigma_{-}+ (t_1+\gamma+\beta^{-1} t_2)\sigma_{+},
\label{eq:nhssh:hamz1}
\end{equation}
where $\sigma_{\pm}=(\sigma_x\pm i\sigma_y)/2$. 
The right and left eigenvectors of the non-Bloch Hamiltonian $H(\beta)$ are defined by  
\begin{equation}
	 H(\beta)\ket{u_\text{R}}=E(\beta) \ket{u_\text{R}},\quad H^\dag(\beta)\ket{u_\text{L}}=E^*(\beta)\ket{u_\text{L}}. 
\end{equation} 
The chiral symmetry guarantees that $\ket{\tilde{u}_\text{R(L)}}\equiv \sigma_z \ket{u_\text{R(L)}}$ is also the right(left) eigenvector of the non-Bloch Hamiltonian, with the corresponding eigenvalues of $-E$ $(-E^*)$.
The right and left eigenvectors satisfy the biorthogonal condition: $\braket{u_\text{L}}{u_\text{R}}= \braket{\tilde{u}_\text{L}}{\tilde{u}_\text{R}}=1, \braket{u_\text{L}}{\tilde{u}_\text{R}}=\braket{\tilde{u}_\text{L}}{u_\text{R}}=0$. 
The projector onto the filled non-Bloch states is defined as $P(\beta) = \ket{\tilde{u}_\text{R}(\beta)} \bra{\tilde{u}_\text{L}(\beta)} $.
The $Q$ matrix for the Hamiltonian is given by the following expression:
\begin{align}
Q(\beta)&=1-2P(\beta), \nonumber \\
&=\ket{u_\text{R}(\beta)}\bra{u_\text{L}(\beta)}-\ket{\tilde{u}_\text{R}(\beta)}\bra{\tilde{u}_\text{L}(\beta)}, \label{Q} 
\end{align}
with $Q^{2}(\beta)=1$. $Q(\beta)$ is a off-diagonal matrix, which can be represented as $Q=\begin{pmatrix}
 	& q  \\
 	q^{-1}  &
 \end{pmatrix}$, due to the chiral symmetry $\sigma_z Q\sigma_z = -Q$. 
Then, the NBWNs can be precisely defined as:
\begin{align}
W &=\frac{1}{4 \pi i} \int_{C_{\beta}} \operatorname{Tr}\left[\sigma_z Q(\beta) d Q(\beta)\right],\nonumber\\
&=\frac{i}{2\pi}\int_{C_{\beta}} q^{-1}dq.\label{winding}
\end{align}

In order to offer a more general illustration, a nonzero $t_3$ case is introduced~\cite{zhongwangPhysRevLett2018a}, whose numerical results are shown in figure~\ref{fig:nhssh_gbz}. 
In this case, GBZ is not a circle which is shown in figure~\ref{fig:nhssh_gbz}(b). 
This implies that the bulk eigenstates with different energies exhibit distinct $|\beta|$.
The energy spectra and the corresponding NBWNs are depicted in figure~\ref{fig:nhssh_gbz}(a), in which $2W$ correctly predicts the quantity of topological zero modes.

It's worth noting that equation~(\ref{winding}) can be extended to multi-band systems.
Every set of bands, denoted by $l$, has a corresponding $C^{(l)}_{\beta}$ curve. The $Q$ matrix in equation~(\ref{Q}) is then represented by $Q^{(l)}$, each of which determines a winding number $W^{(l)}$.
The total winding number, which can be served as a topological invariant, is given by the sum of individual winding numbers: $W=\sum_l W^{(l)}$.
\subsection{Non-Hermitian topological invariants in real space}\label{sec:2.3}
\begin{figure}
\centering
\includegraphics[width=7.6cm, height=5.2cm]{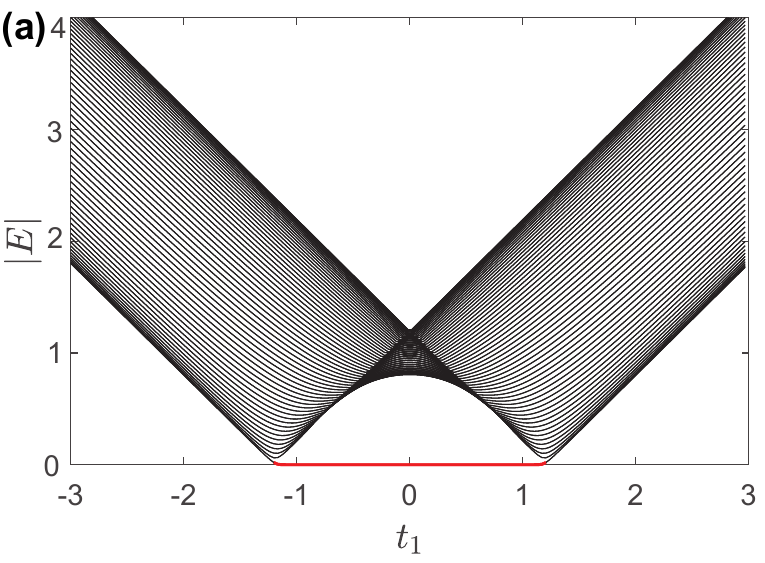}
\includegraphics[width=8cm, height=4.5cm]{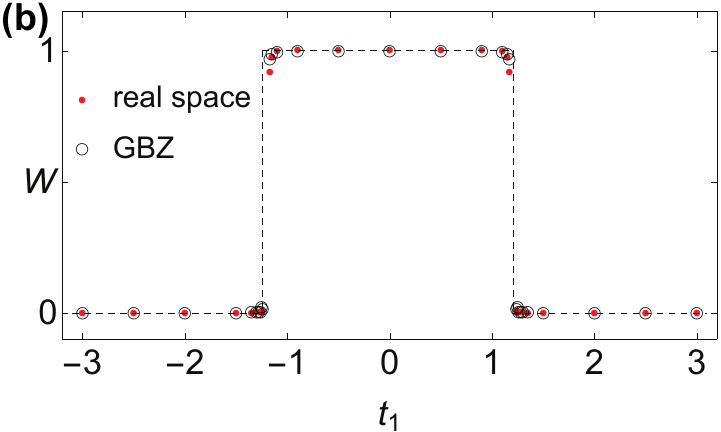}
\caption{(a) The modulus of energy spectrum  for a open chain(L=80) varies with $t_1$. Other parameters are $t_2=1$, $t_3=0.2$, $\gamma=0.1$.  
(b) The red solid dots depict the open-bulk winding number computed in real space with the length of an open chain $L=100$.
 The boundary is cutted at $l=15$. 
 The black hollow dots denote the NBWNs based on the GBZ.
 Reprinted with permission from \cite{wangPRLReal}, Copyright (Year) by the American Physical Society.}
\label{fig:NHSSH_3}
\end{figure}
The non-Hermitian topological invariants can also be formulated directly in real space~\cite{wangPRLReal}, offering a straightforward and comprehensive method to determine non-Hermitian topology.
For the non-Hermitian SSH model with $t_1^L=t_1^R=t_1$, $t_2^L=t_2^R=t_2$, $t_3^L=t_3+\gamma$, $t_3^R=t_3-\gamma$,
The real-space Hamiltonian satisfies $S^{-1}HS=-H$ with $S_{xs,ys'}=\delta_{xy}(\sigma_z)_{s,s'}$. 
Here, $x,y$ represent the index of the unit cell, and $s,s' =A,B$ denotes the degree in sublattice.
The right and left eigenvectors of the real-space Hamiltonian are defined as $H\ket{n_R}=E_n\ket{n_R}$ and $H^\dagger \ket{n_L}=E_n^*\ket{n_L}$.
Due to the chiral symmetry, the energies appear in pairs $(E,-E)$, which lead the eigenvectors defined above correspond to right and left eigenvectors $\ket{\tilde n_R}, \ket{\tilde n_L}$ for $-E$ and $-E^*$.
These eigenvectors are biorthogonal:
$\braket{m_L}{n_R}=\braket{\tilde{m}_L}{\tilde{n}_R}=\delta_{mn},\braket{m_L}{\tilde{n}_R}= \braket{\tilde{m}_L}{n_R} =0$.

From the biorthogonal left and right eigenvectors, the $Q$ matrix was defined via  
\begin{equation}
	Q=\sum_{n} \left(\ket{n_R}\bra{n_L}-\ket{\tilde{n}_R}\bra{\tilde{n}_L}\right),
\end{equation}
where $\sum_n$ represents the summation of all eigenstates within the continuous spectrum of the bulk, excluding the topological edge modes.
The open-bulk winding number~\cite{wangPRLReal} is given by
\begin{equation}
W'= \frac{1}{2L'} \Tr '  (SQ[Q, X]),  \label{open}
\end{equation}
where $X$ is the coordinate operator defined as $X_{xs,ys'}=x\delta_{xy}\delta_{ss'}$, and $\Tr '$ denotes the trace across the middle interval of length $L'$.
The overall length of the chain is given by $L=L'+2l$, in which the interval of boundaries ($1\leq x\leq l$ and $l+L'+1\leq x\leq L$) are not included in the trace. 
The only considered case is the long open chain whose ideal situation is $L\rightarrow \infty$. More specifically, equation~(\ref{open}) for the current model reads $W'= \frac{1}{2L'} \sum_{x=l+1}^{L-l}\sum_{s=A,B} (SQ[Q,X])_{xs,xs}$, where $l$ is large enouth to make sure only the information of bulk remains.
The numerical results of energy spectrum and OPWN under OBC are shown in figure \ref{fig:NHSSH_3}(a) and figure \ref{fig:NHSSH_3}(b), respectively.
The non-zero open-bulk winding number implies the presence of topological modes, revealing the generalized BBC in the non-Hermitian system.

Equation~(\ref{open}) specifically pertains to the topological edge modes exclusively when $\ket{n_R}$ and $\ket{n_L}$ are derived from the OBC, which is significantly different from the Hermitian scenarios where the boundary condition is unimportant.
If the PBC is used instead, the correlation is often lost in the presence of the NHSE.  
The bulk winding number in equation~(\ref{open})  is referred to as an open-bulk topological invariant to stress this special non-Hermitian property.

The Fourier transformation connects the BZ and real space for the common Hermitian bands, making it possible to convert Brillouin-zone topological invariants into real-space topological invariants.
In non-Hermitian systems under the OBC, real space is dual to the GBZ, which indicates that the real-space topological invariant is dual to the GBZ formulation [as depicted in figure \ref{fig:NHSSH_3}(b)].

The duality between the open-bulk winding number and the GBZ formula is not an accident.
A generalized ``Fourier transformation" can be created from the $Q(x)$ matrix~\cite{wangPRLReal}:
\begin{equation}
\tilde{Q}(\beta)=\sum_x Q(x) \beta^{-x},
\end{equation}
which established a connection between $\tilde{Q}(\beta)$ and the $Q(\beta)$.
Indeed,  the series  $\tilde{Q}(\beta)$ converges within a specific region of the complex $\beta$ plane. The curve of GBZ $C_{\beta}$ can be transformed into a curve $\tilde C_{\beta}$ in this region without encountering any singularities of $Q(\beta)$, such as the zero or divergence. 
The following relation was confirmed in reference~\cite{wangPRLReal}:
\begin{equation}
	Q(\beta)|_{\beta\in\tilde C_{\beta}} =\tilde{Q}(\beta)|_{\tilde C_{\beta}},  \label{strong}
\end{equation}
 from which it can be obtained that $Q(\beta)|_{\tilde C_{\beta}}=\sum_x Q(x)\beta^{-x}$. 
 Combined with the NBWNs~\cite{wangPRLReal}, it can be deduced that 
\begin{align}
W =& -\sum_{x,y} \int_{\tilde C_{\beta}} \frac{d\beta}{4\pi i} \Tr[\sigma_z Q(x)\beta^{-x} y Q(y)\beta^{-y-1} ] \nonumber \\
=&\frac{1}{2}\sum_x\text{Tr}[\sigma_z Q(x)xQ(-x)],
\end{align}
in which $\int_{\tilde C_{\beta}} \beta^{-x-y-1}d\beta=2\pi i\delta_{y,-x}$ has been used. 
Here, $xQ(-x) =[Q,X]_{y,y+x}$ is not reliant on $y$ in the bulk. 
Hence,  the NBWN [equation~(\ref{winding})] and open-bulk  winding number [equation~(\ref{open})] can be equivalent by the relation 
\begin{equation}
	W=W'.
\end{equation}
\section{ Periodically driven non-Hermitian systems}\label{sec:3}
\subsection{Floquet theorey}\label{sec:3.1}
Before introducing the periodically driven non-Hermitian system, let us take a brief recall of the Floquet theory~\cite{barone1977floquet,kuchment1982floquet, oka2019floquet,rudner2020band, rudner2020floquet}, which is the basic theory to study the periodically driven systems.
A periodically driven system can be described through a time-periodic Hamiltonian denoted as $H(t)=H(t+T)$, where $T$ represents the time period. 
The evolution of this system can be expanded in a complete and orthonormal basis $\ket{\psi_n(t)}$ called Floquet states, which are defined by $H\ket{\psi}=E\ket{\psi}$ with $\ket{\psi_n(t+T)}=e^{-i\epsilon_n T} \ket{\psi_n(t)}$. 

Bloch's theorem deals with systems that have periodic spatially periodic structures, while Floquet's theorem is for systems that are periodic in time. Similar to Bloch states which can be decomposed into the product of plane waves and periodic functions, Floquet states can be decomposed as:
\begin{equation}
	\ket{\psi_n(t)}=e^{-i\epsilon_n t} \ket{\Phi_n(t)},\label{eq:FloqurtTheorem}
\end{equation}
where $\ket{\Phi_n(t+T)}=\ket{\Phi_n(t)}$ is a periodic function called Floquet modes with the same driving period as the system. 
Thus the time evolution equation became:
\begin{equation}
	(\epsilon_n+i \frac{d }{d t})\ket{\Phi(t)}=H(t)\ket{\Phi(t)}.\label{eq:SE_floquet}
\end{equation}
Since the Hamiltonian $H(t)$ and Floquet modes are both periodic with period $T$. 
Apply the Fourier transition to $\ket{\phi_n} $ and $H(t)$ in frequency domain to obtain
\begin{align}
		\ket{\Phi_n(t)}&=\sum_m e^{im\omega t}\ket{\phi_n^{(m)}}, 
		\nonumber\\
	H(t)&=\sum_m e^{im\omega t} H_{m} . \label{eq:FloquetFourier}
\end{align}
Here $\ket{\phi_n^{(m)}}=\frac{1}{T}\int_{0}^T d t e^{-im\omega t}\ket{\phi_n(t)}$ and $H_{m}=\frac{1}{T}\int_{0}^{T} d t e^{-im\omega t}H(t)$. 
It should be noted that the Fourier coefficients are not normalized and lack a straightforward orthogonality relation. Therefore, in frequency domain, the Schr\"{o}dinger equation  yields:
\begin{equation}
	(\epsilon+m\omega)\ket{\phi_n^{(m)}}=\sum_{m'}H_{m-m'}\ket{\phi_n^{(m')}}.\label{eq:SE_floquet_omega}
\end{equation}

The quasienergy spectrum can be obtained by extending the Hilbert space with bases $\ket{\varphi_n}=\nket{\phi_n^{(m)}}\otimes\ket{m}$ where $m$ is called the Fourier index. 
Then equation~(\ref{eq:SE_floquet_omega}) can be expressed as an eigenvalue equation in Fourier harmonic space: $\mathcal H \varphi_n=\epsilon_{n} \varphi_n$, where $\mathcal H$ is the Floquet Hamiltonian whose matrix elements are: 
\begin{equation}
	\bra{\varphi^{m'}_{n'}}\mathcal H\ket{\varphi^{m}_n}=\bra{\psi_{n'}}H_{m'-m}\ket{\psi_{n}}- \delta_{n,n'}\delta_{m,m'}m\omega.
\end{equation}

For more detailed, the matrix form of $\mathcal{H}$ and $\ket{\varphi_n}$ take the following formula:
\begin{equation*}
		\mathcal{H}=
		\begin{pmatrix}
		\ddots & H_{-1} & H_{-2} & \\
		H_{1} & H_0-m  \omega & H_{-1} & H_{-2} \\
		H_{2} & H_{1} & H_0-(m+1) \omega & H_{-1} \\
		& H_{2} & H_{1} & \ddots
	\end{pmatrix},
\end{equation*}
\begin{equation}
	\ket{\varphi_n}=\begin{pmatrix}
		\cdots &
		\ket{\phi_n^{(m)}} &
		 \ket{\phi_n^{(m+1)}} &
		\cdots
	\end{pmatrix}^T.
\end{equation}

Note that the Floquet matrix has a block structure where the size of each block is the same as the system. 
The number of blocks labeled by Floquet numbers is infinite. 
The time-periodic component of the Floquet state wave function, denoted as $\ket{\Phi(t)}$, is derived by multiplying with a rectangular matrix of oscillatory phase factors, $\mathcal P(\omega t) =(\cdots e^{-im\omega t} e^{-i(m+1)\omega t} \cdots)$:
\begin{align}
	\ket{\Phi_n(t)}=\mathcal{P} (\omega t) \bm{\varphi}_n=\sum_m e^{-im\omega t}\ket{\varphi_n^{(m)}}.
\end{align}

From equation~(\ref{eq:FloqurtTheorem}) and equation~(\ref{eq:FloquetFourier}), the Floquet states can be expressed as:
\begin{align}
	\ket{\psi_n}&=e^{-i\epsilon_n t}\sum_m e^{-im\omega t}\ket{\phi_n^{(m)}}
	\nonumber\\
	&=e^{-i(\epsilon_n+m'+\omega )t}\sum_m e^{-i(m-m')\omega t}\ket{\phi_n^{(m)}}
	\nonumber\\
	&=e^{-i\tilde\epsilon_n t}\sum_m e^{-im\omega t}\ket{\tilde\phi_n^{(m)}},
\end{align}
in which $\tilde\epsilon_n=\epsilon_n+m' \omega$ and $\nket{\tilde\phi_n^{(m)}}=\nket{\phi_n^{(m+m')}}$. 
Therefore, by introducing integer multiples of $\omega$ in the pair of exponentials in this expression, the quasienergies can be shifted.
This implies that all unique solutions to the Schr\"odinger equation, known as Floquet states, can be categorized using quasienergies that fall within a specific range called the "Floquet-Brillouin zone"~\cite{rudner2020floquet}.
The width of this zone is $\omega$, meaning that the quasienergies range from $\epsilon_{\text{min}}$ to $\epsilon_{\text{min}}+\omega$. 
Consequently, the Floquet Hamiltonian's spectrum will exhibit a periodic structure, comprising an infinite number of copies of the system's Floquet spectrum within the Floquet-Brillouin zone. 

The time-evolution operator is 
\begin{equation}
	U(k,t)= \mathcal{T} \mathrm{exp}\left[-i \int_{0}^{t} dt' H(k,t')\right],
\end{equation}
in which $\mathcal{T}$ represents time-ordering.
Floquet states are stationary states of the stroboscopic evolution operator (or named Floquet operator)
\begin{equation}
	U(T)=\mathcal T \exp\left[-i\int_{0}^T d t'H(t')\right].
\end{equation}
The Floquet operator propagates the system forward in time through one complete period of the drive.
From the Floquet operator, the Floquet effective Hamiltonian can be defined as
\begin{equation}
	H_{\text{eff}}=\frac{i}{T}\ln U.
\end{equation}

\subsection{Harmonically driven non-Hermitian SSH model}\label{sec:3.2}
In a static non-Hermitian system with the NHSE, the BBC based on the Bloch band theory should be extended to the generalized BBC based on the non-Bloch band theory in which the GBZ takes an essential place.
The NHSE can emerge in a periodically driven non-Hermitian system, in which the topological invariants based on BZ cannot accurately characterize the edge states.
Then, the non-Hermitian topological invariants should be defined on the GBZ instead of the BZ  to establish the generalized BBC for the periodically driven non-Hermitian systems. 
Here we review the non-Bloch band theory in the 1D harmonically driven non-Hermitian SSH model.
\subsubsection{Model}
\begin{figure}
	\centering
	\includegraphics[width=0.45\textwidth]{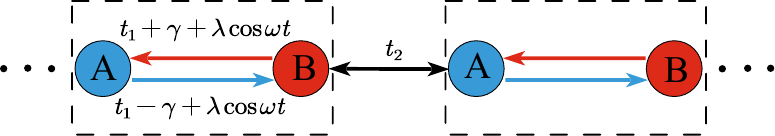}
	\caption{The sketch of a harmonically driven non-Hermitian SSH model.}
	\label{fig:harmo_model}
\end{figure}
~The 1D harmonically driven non-Hermitian SSH model can be constructed by introducing periodic driving to the intracell hopping~\cite{cao2021non}.
For simplicity, only the nearest neighbor hopping is considered and the nonreciprocity and harmonically periodic driving are introduced into intracell hopping.
A sketch of this model is depicted in figure~\ref{fig:harmo_model}, and its Hamiltonian in real space is 
	\begin{align}
		H=&\sum_n [t_1+\gamma+\lambda\cos (\omega t)]c_{nA}^{\dagger}c_{nB}^{} \nonumber
		\\ &+[t_1-\gamma+\lambda\cos (\omega t)]c_{nB}^{\dagger}c_{nA}^{} 
		\nonumber \\
		&+t_2c_{nB}^{\dagger}c_{n+1A}^{}+t_2c_{n+1A}^{\dagger}c_{nB}^{}.\label{eq:harmo_hamr}
	\end{align}
	Here, $\lambda$ represents the strength of periodic driving and $\gamma$ represents the strength of non-Hermiticity.
	The Bloch Hamiltonian has the following formula:
	\begin{align}
	H(k,t) &= [n_{x} + \lambda \cos(\omega t)]\sigma_{x} + (n_{y} +i \gamma)\sigma_{y},
	\nonumber\\
	n_{x} &= t_{1} + t_{2} \cos(k),\quad
	 n_{y} = t_{2} \sin(k).
	\label{eq:harmo_hamk}
	\end{align}
where $\sigma_{x,y}$  represent Pauli matrices.
\begin{figure*}
	\centering
	\includegraphics[width=0.95\textwidth]{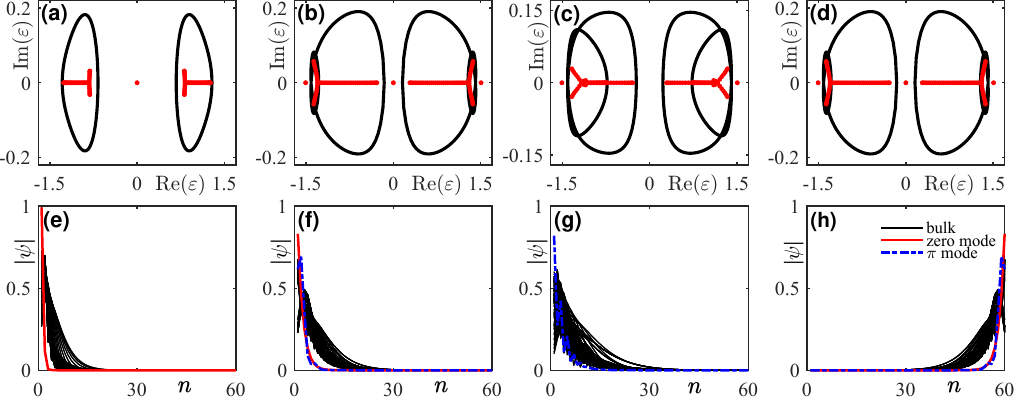}
	\caption{Energy spectra and the profile of eigenstates for the non-Hermitian SSH model with harmonic driving. (a)--(d): The quasienergy spectra of the four distinct topological phases are shown under two different boundary conditions: PBC in black and OBC in red. (e)--(h) The density distributions of the bulk states (black solided), zero modes (red solided), and $\pi$ modes (blue dotted) of the four distinct phases in (a)--(d), respectively. $n$ represents the unit cell's index. Parameters are $t_2=1$, $\gamma$=0.2, $\lambda=0.5$ $\omega=3$, (a)(e) $t_{1} = 0.3$, (b)(f) $t_{1} = 0.75$, (c)(f) $t_{1} = 1.3$, and (d)(h) $t_{1} = -0.75$. Reprinted with permission from ~\cite{cao2021non}, Copyright (Year) by the American Physical Society.} \label{se}
\end{figure*}
The Bloch Hamiltonian exhibits periodicity in time $H(k,t) = H(k,t+T)$ with time period $T=2 \pi/\omega$.
Additionally, it possesses a chiral symmetry given by $\sigma_{z} H(k,t) \sigma_{z}= - H(k, -t)$, which guarantees that quasienergies occur in pairs of $(E, -E)$.
The Floquet state of band $n$ evolves according to the Schr\"odinger equation:
\begin{equation}
i\partial_{t} \ket{\psi_{n,R}(k,t)} = H(k,t) \ket{\psi_{n,R}(k,t)}.
\end{equation}
The Floquet states can be derived using Fourier transformation and Floquet theorem and take the form of
\begin{align}
	\nket{\psi_{n,R}(k,t)} = e^{- i \varepsilon_{n}(k) t }\sum_{m} e^{i m \omega t } \nket{\psi_{n,R}^{(m)}(k)},
\end{align}
in which $\varepsilon_{n} $ represents the quasienergy and $\nket{\psi_{n,R}^{(m)}(k)}$ denotes a corresponding right eigenvector. 
The formulation of the Schr\"odinger equation in the frequency domain is 
\begin{equation}
\sum_{m'} \mathcal{H}_{m,m'}(k) \ket{\psi_{n,R}^{(m')}(k)} = \varepsilon_{n}(k)\ket{\psi_{n,R}^{(m)}(k)}. \label{floqh2}
\end{equation}
Here $\mathcal{H}_{m,m'}(k) = m \omega \delta_{m,m'} \mathbf{I} + H_{m-m'}(k)$ and $H_{m}(k) = \frac{1}{T} \int_{0}^{T}dt H(k,t) \mathrm{exp}(-i m\omega t)$ are the Floquet Hamiltonian. 
More explicitly, the Floquet Hamiltonian takes the following form
\begin{equation}
\mathcal{H}=
\left(
  \begin{array}{ccccc}
    ... &        &      &      &   \\
        & H_{0}+ \omega & H_{1} & 0 &   \\
        &  H_{-1} & H_{0} & H_{1}  &  \\
        &  0 & H_{-1}  & H_{0} - \omega &  \\
        &   &   &    &   ... \\
  \end{array}
\right),
\label{eq:harmo_hamf}
\end{equation}
in which $H_{0}=n_{x}\sigma_{x} + [n_{y} +i \gamma]\sigma_{y}$ and $H_{\pm 1} = {\lambda} \sigma_{x}/2$. 
The quasienergies, which are the eigenvalues of this Floquet Hamiltonian, are coupled in the form of $(E,-E)$ due to the presence of chiral symmetry $\mathcal{C}^{-1} \mathcal{H} \mathcal{C} = -\mathcal{H}$~\cite{cao2021non}.

\subsubsection{Non-Hermitian skin effect} 
~Due to the infinite rank of the Floquet Hamiltonian $\mathcal{H}(k)$, a truncation should be made to obtain the quasienergies.
Figures~\ref{se}(a)--(d) display the quasienergy spectra for four distinct phases based on the Floquet Hamiltonian under OBC (marked in red) and PBC (marked in black), respectively.
Periodic driving can cause an unusual gap to emerge at the quasienergy $\epsilon=\pi$, allowing for the stability of topological $\pi$ modes. 
Therefore, two distinct types of topological edge modes can occur in the two gaps with $\epsilon=0,\pi$. 
The quasienergy spectra exhibit significant distinction between the periodically driven systems under OBC and PBC, indicating the emergence of the NHSE in the periodically driven non-Hermitian SSH under the OBC. 
The wave functions of bulk states and the topological edge states correspond to figures~\ref{se}(a)--(d) under the OBC are depicted in figure~\ref{se}(e)--(h), in which the black, red, and blue line denotes the bulk states, zero modes, and $\pi$ modes, respectively.
As shown in figures~\ref{se}(a) and (e), only the topological zero modes emerge at the left end of the open chain. 
The topological phase only has the $\pi$ modes are shown in figures~\ref{se}(c) and (g). 
The phases have both the topological zero modes and $\pi$ modes are shown in figures~\ref{se}(b) (f) and figures~\ref{se}(d) (h). 
The bulk states are localized at the left end of the open chain and exhibit the NHSE  in figure~\ref{se}(e)--(g). When changing the sign of $t_{1}$, the bulk states will accumulate at opposite ends of the open chain as shown in figure~\ref{se}(h). 
As previously stated, the Bloch theory is no longer applicable in the non-Hermitian systems with the NHSE, and the Bloch BBC is broken down.
The Bloch band theory needs to be generalized into the non-Bloch band theory, which is founded on the concept of GBZ. 

\begin{figure}
	\centering
	\includegraphics[width=0.45\textwidth]{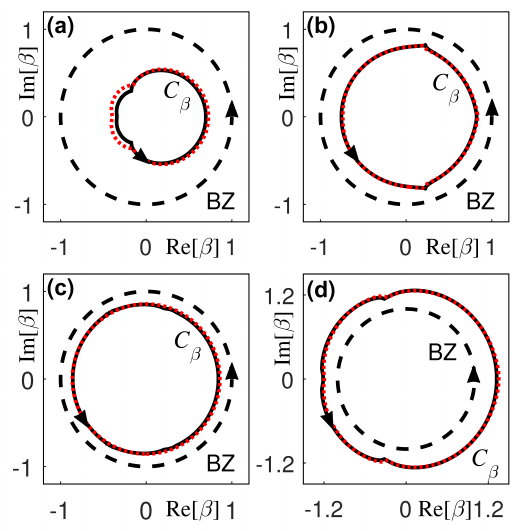}
	\caption{The GBZ of the four different topological phases in figures~\ref{se}(a)--(d)~\cite{cao2021non}, obtained by the non-Bloch effective Hamiltonian (dotted curves) and non-Bloch Floquet Hamiltonian (solid curvess), respectively. The BZ is denoted by a dashed unit circle. } \label{fig:harmo_gbz}
\end{figure}

\subsubsection{Non-Bloch winding numbers}
The non-Bloch band theory is essential for the accurate description of the non-Hermitian systems, where the GBZ plays a significant role.
To establish the GBZ, the Bloch Floquet Hamiltonian $\mathcal{H}(k)$ should be rewrited  by $e^{ik}\rightarrow \beta=re^{ik}, k\in \mathbb{R}$ as non-Bloch Floquet Hamiltonian $\mathcal{H}(\beta)$, which has the chiral symmetry with $\sigma_{z} H(\beta,t) \sigma_{z} = -H(\beta, -t)$. 
Here, $\beta$ represents the solutions of the characteristic equation $\det[\mathcal{H}(\beta) - E]=0$, which is a polynomial equation of $\beta$ with an even degree. 
The quasienergies are restricted to a single period, especially $\epsilon \in [-\pi, \pi]$, due to $2\pi$ modules of the quasienergy $\epsilon$. 
The solutions of $\beta$, whose total number is $2N$, can be sorted as $|\beta_{1}| \leq |\beta_{2}| \leq ... \leq |\beta_{2N}|$. 
Only the two solutions satisfied the continuum band condition $|\beta_{N}|=|\beta_{N+1}|$ belongs to the GBZ~\cite{zhongwangPhysRevLett2018a,yokomizo2019non,cao2021non}.
The GBZ can be captured by $\beta_{N}$ and $\beta_{N+1}$ in the complex plane ($C_{\beta}$), as shown in figures~\ref{fig:harmo_gbz}(a)--(d) for four phases in figure~\ref{se} with $t_{1}=0.3$, $0.75$, $1.3$ and $-0.75$, respectively. 
The distinction between the GBZ and the BZ leads to the quasienergies' collapse and the NHSE.
In figures~\ref{fig:harmo_gbz}(a)--(c), the GBZ is smaller than BZ which results in the localization of bulk eigenstates at the left end of the open chain. 
The GBZ is larger than BZ in figure~\ref{fig:harmo_gbz}(d), which causes the bulk eigenstates to be localized at the right end of the open chain.
    
The non-Bloch time-evolution operator can be derived from the non-Bloch Hamiltonian $H(\beta,t)$ with the formula of
\begin{equation}
U(\beta,t)= \mathcal{T} \mathrm{exp}\left[-i \int_{0}^{t} dt' H(\beta,t')\right],
\end{equation}
in which $\mathcal{T}$ represents time-ordering.
It  exhibits the symmetry with $\sigma_{z} U(\beta,t) \sigma_{z}=U ^{-1}(\beta, t) = U (\beta, - t)$ and obeys the equation as following:
\begin{equation}
i\partial_{t} U(\beta,t) = H(\beta,t) U(\beta,t).
\end{equation}

When the eigenenergies of the above non-Bloch Floquet operator has a gap at $e^{-i \epsilon}$,  a well-defined non-Bloch effective Hamiltonian at the gap takes the following formula:
\begin{equation}
H_{\mathrm{eff}}^{\epsilon} (\beta) = \frac{i}{T} \mathrm{ln}_{- \epsilon} U(\beta,T),\label{enbh}
\end{equation}
where  $\epsilon$ denotes the branch cut. Here, we choose $\ln_{\epsilon} e^{i \phi} = i \phi$ defined at the interval $\epsilon - 2\pi < \phi < \epsilon$. The  effective Hamiltonians can be written in the terms of left and right eigenvectors with $H_{\mathrm{eff}}^{\epsilon} (\beta) =  \frac{i}{T} \sum _{n} \mathrm{ln}_{- \epsilon} \left( \lambda_{n}(\beta) \right) | \psi_{n,R}(\beta) \rangle \langle \psi_{n,L}(\beta) |$.
The chiral symmetry is broken:
\begin{align}
    & \sigma_{z}H_{\mathrm{eff}}^{\epsilon} (\beta) \sigma_{z}\nonumber\\
    =&\frac{i}{T}  \sigma_{z} \left[ \ln_{-\epsilon}  U (\beta,  T) \right] \sigma_{z}, \nonumber\\
    =&\frac{i}{T} \ln_{-\epsilon} \left[ \sigma_{z} U (\beta,  T) \sigma_{z} \right], \nonumber\\
    = &\frac{i}{T} \ln_{-\epsilon}  U ^{-1} (\beta,  T),  \nonumber\\
    = &\frac{i}{T} \sum_{n} \ln_{-\epsilon} \left( \lambda_{n}^{-1}\right) | \psi_{n,R}(\beta) \rangle \langle \psi_{n,L}(\beta) |,  \nonumber\\
    =&- \frac{i}{T} \sum_{n}  \left[  \ln_{-\epsilon} (\lambda_{n}) + 2\pi i \right] | \psi_{n,R}(\beta) \rangle \langle \psi_{n,L}(\beta) |,  \nonumber\\
    =&- \frac{i}{T} \sum_{n}  \ln_{-\epsilon} (\lambda_{n})  | \psi_{n,R}(\beta) \rangle \langle \psi_{n,L}(\beta) | +  \omega,  \nonumber\\
    = &- H_{\mathrm{eff}}^{- \epsilon} (\beta) + \omega.
\end{align}

During each period, the effective Hamiltonians solely reflect the stroboscopic dynamics with respect to quasienergies $\epsilon$ while losing the important informations of the time evolution. 
As a result, the topological invariants should be based on the NBPEOs, which is given as:
\begin{equation}
U_{\epsilon}(\beta,t)= U(\beta,t) e^{i  H_{\mathrm{eff}}^{\epsilon} (\beta) t }.
\end{equation}
The NBPEOs are periodic in time with $U_{\epsilon}(\beta,t + T) = U_{\epsilon}(\beta,t)$ and have the symmetry:
\begin{align}
    &\sigma_{z} U_{\epsilon} (\beta,t) \sigma_{z},\nonumber\\
    = &\sigma_{z} U (\beta,t) \sigma_{z}^{2}  e^{ i t H_{\mathrm{eff}}^{\epsilon} (\beta)} \sigma_{z},\nonumber\\
    =&  U (\beta, -t) e^{i t \sigma_{z} H_{\mathrm{eff}}^{\epsilon} (\beta) \sigma_{z}}, \nonumber\\
    = &U (\beta, -t) e^{- i  t  H_{\mathrm{eff}}^{-\epsilon} (\beta)  + i \omega t}, \nonumber\\
    =& U_{-\epsilon} (\beta, -t) e^{i \omega t} .
\end{align}
Then, the NBPEOs exhibit chiral symmetry during a half-period, with quasienergies $\epsilon=0$ and $\pi$:
\begin{align}
\sigma_{z} U_{0/\pi}\left(\beta,\frac{T}{2}\right) \sigma_{z} &= \mp U_{0/\pi}\left(\beta,\frac{T}{2}\right).
\end{align}
Due to the chiral symmetry, the NBPEO during a half-period  is off-diagonal at $\epsilon=0$ and diagonal at $\epsilon=0$:
\begin{align}
U_{0}\left(\beta,\frac{T}{2}\right)&=
\left(
  \begin{array}{cc}
       0 & U_{0}^{+}(\beta) \\
      U_{0}^{-}(\beta)  &  0 \\
  \end{array}
\right),\\
U_{\pi}\left(\beta,\frac{T}{2}\right)&=
\left(
  \begin{array}{cc}
      U_{\pi}^{+}(\beta)  &  0 \\
       0 &  U_{\pi}^{-}(\beta)  \\
  \end{array}
\right).
\end{align}
By the Yao-Wang formula~\cite{zhongwangPhysRevLett2018a, XiaoLeiNaturePhysics2020}, NBWNs can be defined for harmonically driven non-Hermitian SSH model~\cite{cao2021non} based on the GBZ as:
\begin{equation}
W_{\epsilon=0,\pi} = \frac{i}{2 \pi} \int_{C_{\beta}} \mathrm{Tr}\left[\left(U_{\epsilon}^{+}(\beta)\right)^{-1} d U_{\epsilon}^{+}(\beta)\right].
\label{nbwn}
\end{equation}
Here, $W_{0}$ and $W_{\pi}$ characterize the topological zero modes and $\pi$ modes, respectively. 
\begin{figure}
	\centering
	\includegraphics[width=0.45\textwidth]{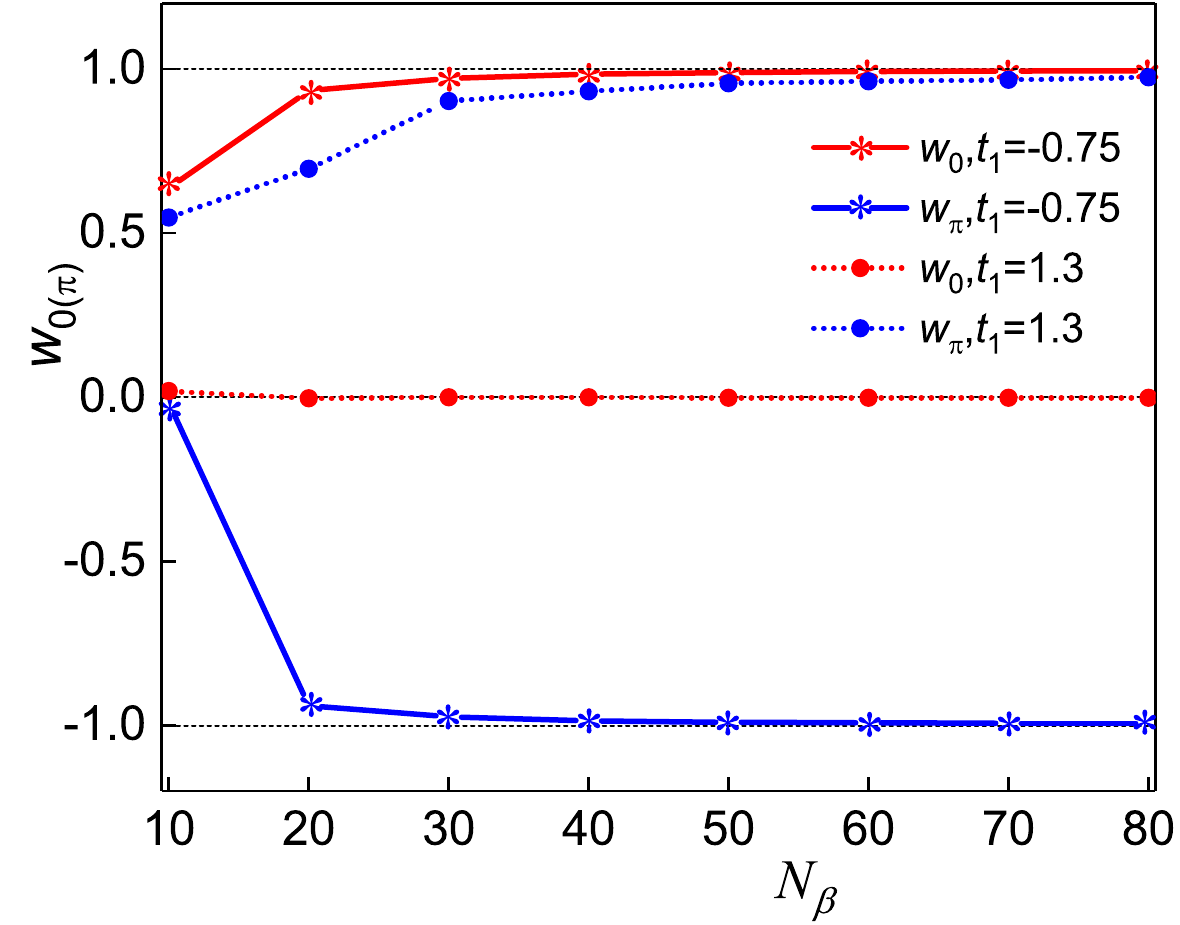}
	\caption{The NBWNs $W_{0/\pi}$ with varied GBZ size $N_{\beta}$ for $t_1=-0.75$(solid lines) and $t_1=1.3$(dotted lines)~\cite{cao2021non}. $W_0$ and $W_\pi$ are marked with red and blue, respectively.} %
	\label{fig:harmo_winding}
\end{figure}

As an illustration, the numerical results of NBWNs for zero modes and $\pi$ modes are shown in figure~\ref{fig:harmo_winding}, in which the values of NBWNs will rapidly converge to an integer as the size of the GBZ increases.
NBWNs are equal to $\pm1$ when the topological zero modes and $\pi$ modes appear, otherwise it is zero.
Therefore, the generalized BBC is established by defining the NBWNs ($W_{0/\pi}$) to characterize the topologically nontrivial zero modes and $\pi$ modes.

\begin{figure}
	\centering
	\includegraphics[width=0.45\textwidth]{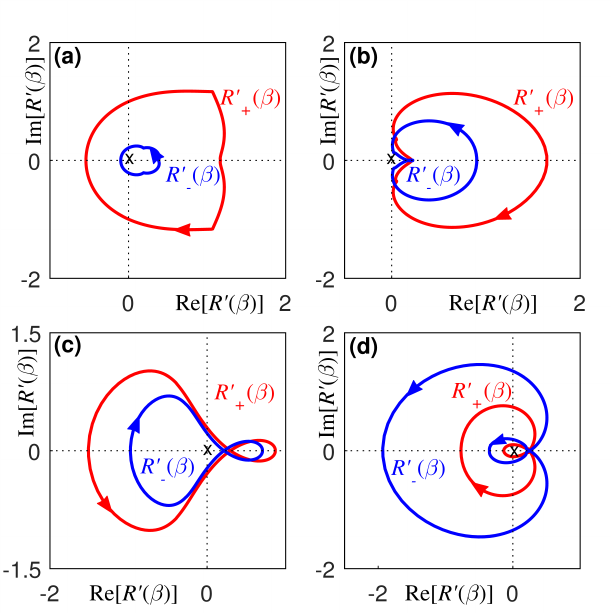}
	\caption{(a)--(d) The $R'_{+} (\beta)$ and $R'_{-} ( \beta)$ loops along the GBZs of figures~\ref{fig:harmo_gbz}(a)--(d) on the complex plane, respectively. Reprinted with permission from~\cite{cao2021non}, Copyright (Year) by the American Physical Society.} \label{fig:harmo_loop}
\end{figure}

\subsubsection{Non-Bloch band invariants}
~Based on the Floquet operator, the NBWNs can be defined to predict the emergence of the Floquet topological edge states. Meanwhile, the topology of the Floquet bands can be defined intuitively by the non-Bloch effective Hamiltonian in terms of the above Floquet operators:
\begin{eqnarray}
    H_{\mathrm{eff}} (\beta) &=& \frac{i}{T} \mathrm{ln} U(\beta,T),\nonumber\\
    &=& H_{0}(\beta) + \frac{[H_{0}(\beta), H_{1}]}{\omega} -\frac{[H_{0}(\beta), H_{-1}]}{\omega} \nonumber\\
    & & + \frac{[H_{1}, H_{-1}]}{\omega}.
\end{eqnarray}

For the case of high-frequency ($\omega$ larger than the bandwidth of $H_{0}(\beta)$), the Floquet bands have no overlap in the Floquet Brillouin zone. Then, the periodic driving does not affect the effective Hamiltonian with: $H_{\mathrm{eff}}(\beta)= H_{0}(\beta)$. 
The GBZ is a circle $C_{\beta} = r e^{i \theta}$ with $\theta \in [0, 2\pi)$ and $r=\sqrt{\abs{(t_{1} -\gamma)/(t_{1} + \gamma)}}$.

For the case of low-frequency ($\omega$ smaller than the bandwidth), a single resonance can be generated by the driving between the two non-Bloch bands of $H_{0}(\beta)$. 
Thus, a non-Bloch effective Hamiltonian at low-frequency case is derived with a rotated Floquet operator 
\begin{equation}
    U(\beta,T) = \mathcal{T} \mathrm{exp}\left[-i \int_{0}^{T} dt~H_{\mathrm{R}}(\beta,t)\right],
\end{equation}
where $H_{\mathrm{R}}(\beta,t)$ is the rotated Hamiltonian obtained by applying a rotation $\hat{O}(\beta,t)$ into the Hamiltonian $H(\beta,t)$.
The rotation operator is 
\begin{equation}
     \hat{O}(\beta,t)= \exp \left[- i \hat{\mathbf{n}}(\beta) \cdot \mathbf{\sigma} \omega t /2 \right],
\end{equation}
with $\hat{\bf{n}}(\beta) = {\bf n}(\beta) / n(\beta)$ and $n(\beta)=\sqrt{n_{x}^{2}(\beta)+n_{y}^{2}(\beta)}$. 
Then, the effective Hamiltonian can be derived as follows:
\begin{equation}
    H_{\mathrm{eff}} (\beta) = \left(
    \begin{array}{cc}
    0 &   R'_{+}(\beta)    \\
    R'_{-}(\beta)  &  0  \\
    \end{array}
    \right),
\end{equation}
with
\begin{eqnarray}
    R'_{\pm}(\beta) = \left[1- \frac{\omega}{2 n_{\beta}} \mp \frac{\lambda[R_{+}(\beta)-R_{-}(\beta)]}{4 n_{\beta}^{2}} \right] R_{\pm}(\beta),
\end{eqnarray}
and $R_{\pm}(\beta) = t_{1} \pm \gamma + t_{2} \beta^{\mp}$.
The GBZ of the effective Hamiltonian can be determined by the solution of the characteristic equation $\det [H_\text{eff}(\beta)-E]=0$.
The GBZ $C_{\beta}$ determined from effective Hamiltonians, shown in figures~\ref{fig:harmo_gbz}(a)--(d) with dotted curves, is consistent with the GBZs (solided curves) determined by the above Floquet Hamiltonian. The generalzied "Q matrix" is off-diagonal:
\begin{align}
Q(\beta) =& 
\left(
\begin{array}{cc}
    0 &   q    \\
    q^{-1}  &  0  \\
\end{array}
\right),
\nonumber\\
=&\frac{1}{\sqrt{R'_{+}(\beta) R'_{-}(\beta)}} [R'_{+}(\beta) \sigma_{+} + R'_{-}(\beta) \sigma_{-}], 
\label{Qmat}
\end{align}
which leads to $q = R'_{+}(\beta)/\sqrt{R'_{+}(\beta) R'_{-}(\beta)}$.
The non-Bloch band invariant can be expressed by Yao-Wang formula as
\begin{align}
\mathcal{W} &= \frac{i}{2 \pi} \int_{C_{\beta}} d q ~ q^{-1}(\beta), \nonumber\\
 &=\frac{1}{4\pi} \left[ \arg R'_{-}(\beta) - \arg R'_{+}(\beta) \right]_{C_{\beta}}.
\label{nbcn}
\end{align}
This indicates that the band invariant can be calculated by the changes in the phases of the $R'_{+}(\beta)$ and $R'_{-}(\beta)$ in the effective Hamiltonian as $\beta$ traverses the counterclockwise path $C_{\beta}$ in GBZ.
The loops of $R'_{+} ( \beta )$ are shown in red curves and the loops of $R'_{-} ( \beta )$ are shown in blue curves in figures ~\ref{fig:harmo_loop}(a)--(d), along the GBZ of figures~\ref{fig:harmo_gbz} (a)--(d).

The non-Bloch band invariants for figures~\ref{fig:harmo_loop}(a)--(d) are $1, 0, -1$, and $2$, respectively.
As shown in figure~\ref{fig:harmo_loop}(b), the original point is not surrounded by both of the $R'_{+} (\beta)$ and $R'_{-} (\beta)$ on the complex plane, implying the trivial topology of the Floquet bands. 
But, the phase possesses robust zero modes and $\pi$ modes and exhibits nontrivial topology with non-zero NBWNs ($W_{0}=W_{\pi}=1$).
Therefore, the band invariant solely highlights the topology of the Floquet bands and does not adequately describe the robust topological zero modes and $\pi$ modes.
The phenomenon occurs due to the loss of time evolution information within each period by the Floquet operator. 
As illustrated in figure~\ref{fig:harmo_loop}(d),  both of the two loops undergo rotations around the origin point, while their directions are opposite.  
This shows that the band invariant is $2$ while the NBWNs are nontrivial with $W_{0}=1$ and $W_{\pi}=-1$, indicating that NBWNs possess greater fundamental significance compared to the band invariant. Indeed, NBWNs and the band invariant are related, and this can be demonstrated as $W_{0} - W_{\pi} = \mathcal{W}$. We can define a difference by the effective Hamiltonians at the two gaps:
\begin{align}
 &H_{\mathrm{eff}}^{\pi} (\beta) -  H_{\mathrm{eff}}^{0} (\beta),\nonumber\\
=& i \mathrm{ln}_{- \pi} U(\beta,T)/T - i \mathrm{ln}_{0} U(\beta,T)/T,
\nonumber\\
=& i \sum_{n} \left[\ln_{-\pi} \left( \lambda_{n}\right) - \ln_{0} \left( \lambda_{n}\right) \right]| \psi_{n,R}(\beta) \rangle \langle \psi_{n,L}(\beta) |/T,
\nonumber\\
=& {i} \sum_{0<\epsilon_{n}<\pi} \left[-i\epsilon_{n}- 2 \pi i + i\epsilon_{n} \right]| \psi_{n,R}(\beta) \rangle \langle \psi_{n,L}(\beta) |/T,
 \nonumber\\
=& {2\pi} \sum_{0<\epsilon_{n}<\pi} | \psi_{n,R}(\beta) \rangle \langle \psi_{n,L}(\beta) |/T,
 \nonumber\\
=&\omega P_{0,\pi}(\beta).
\end{align}
The NBPEOs at half period is:
\begin{align}
&U_\pi ^{-1}\left(\beta, T/2\right) U_0\left(\beta, T/2\right),\nonumber\\
=& e^{  - i  H_{\mathrm{eff}}^{\pi} (\beta) T/2 } U(\beta, -T/2) U(\beta, T/2) e^{ i  H_{\mathrm{eff}}^{0} (\beta) T/2 }, \nonumber\\
=& e^{- i \pi P_{0,\pi}},\nonumber\\
=& 1 - 2 P_{0,\pi},\nonumber\\
=& Q(\beta).
\end{align}
\begin{figure}
	\centering
	\includegraphics[width=0.48\textwidth]{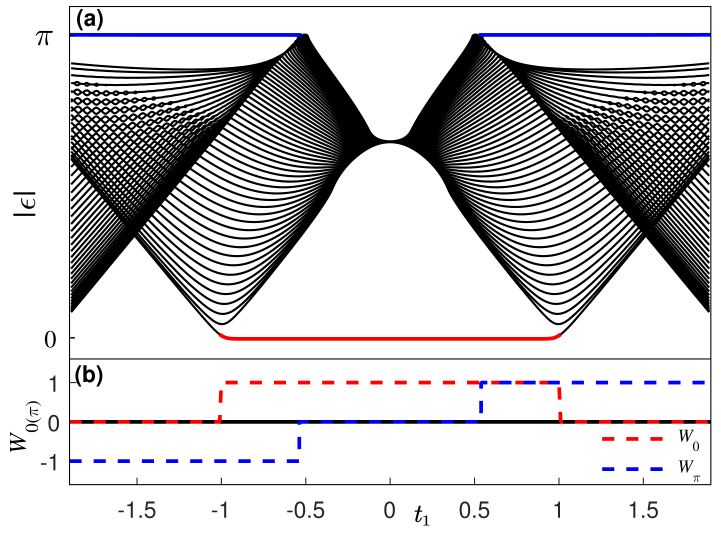}
	\caption{(a) Quasienergy spectrum as functions of $t_1$ for the harmonically driven non-Hermitian SSH model under the OBC.
		(b)The corresponding NBWNs for edge states with quasienergies $\epsilon =0$ (red dashed line) and $\epsilon =\pi$ (blue dashed line). Reprinted with permission from~\cite{cao2021non}, Copyright (Year) by the American Physical Society.
	} \label{fig:harmo_bbc}
\end{figure}

From the property of the NBWNs, the relationship can be demonstrated as follows:
\begin{align}
&W\left[U_\pi ^{-1}\left(\beta, T/2\right) U_0\left(\beta, T/2\right)\right], \nonumber\\ 
=& W\left[U_0 \left(\beta, T/2\right) \right] - W\left[U_\pi \left(\beta, T/2\right) \right], \nonumber\\ 
=&W_0-W_\pi= \mathcal{W}. 
\end{align}

\subsubsection{Topological phase transition}

\begin{figure}
	\centering
    \includegraphics[width=0.4\textwidth]{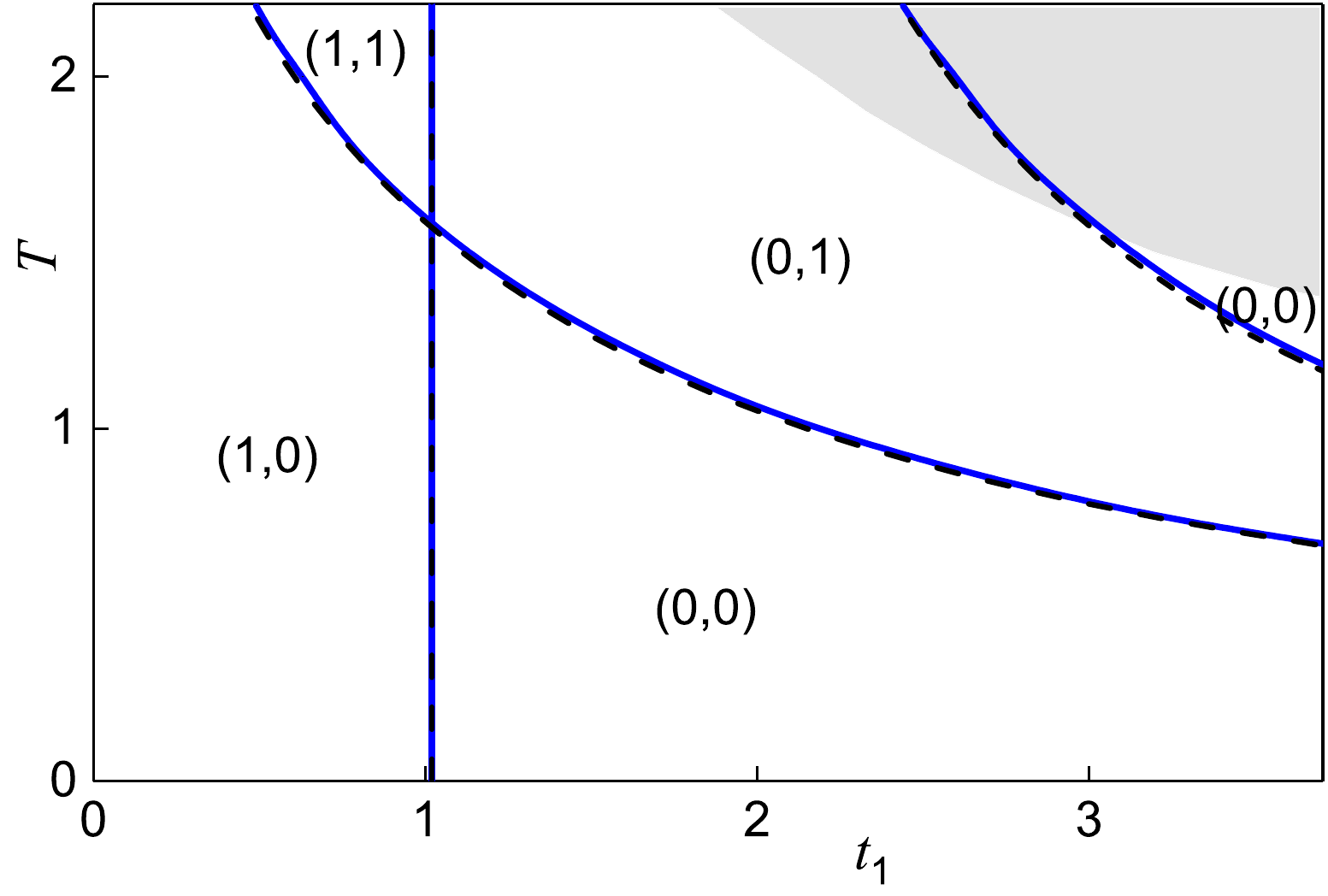}
    \caption{The phase diagram of the Floquet non-Hermitian SSH model in equation~(\ref{eq:harmo_hamr}). 
    	The dashed curves are calculated by the non-Bloch band theory.
    	The solid curves are numerical results of the Floquet Hamiltonian under OBC.
    	Other parameters are  $t_{2}=1$, $\gamma=0.2$ and $\lambda=0.5$.
    	The non-Bloch topological invariants cannot be well defined with gapless quasienergies, which are denoted by the shaded region in the diagram.
    	There exist four different topological phases in the diagram, denoted by the different NBWNs $(W_{0/\pi})$. Reprinted with permission from~\cite{cao2021non}, Copyright (Year) by the American Physical Society.
    	 }  \label{pd}
\end{figure}
~The NBWNs of the harmonically driven non-Hermitian SSH model are defined to predict the Floquet topological edge modes.
The NBWNs provide a powerful tool to investigate the topological phase transitions.
Figure~\ref{fig:harmo_bbc}(a) shows dimensionless quasienergies $\epsilon=\varepsilon T$ of Floquet Hamiltonian varies with $t_1$, in which two different types of Floquet edge states can be observed: zero modes and $\pi$ modes, marked by red and blue colors, respectively.
The corresponding NBWNs $W_0$ and $W_{\pi}$ are shown in figure~\ref{fig:harmo_bbc}(b) with red and blue dashed lines, respectively.

The phase doundaries can be determined by the effective Hamiltonian. In the scenario of week driving, the expression of $R'_{\pm}$ is able to be simplified as follows:
\begin{equation}
R'_{\pm}(\beta) \simeq \left[1- \frac{\omega}{2 n_{\beta}} \right] R_{\pm}(\beta).
\end{equation}
Then, the GBZ can be determined and written as
\begin{equation}
C_{\beta} = r e^{i \theta},
\end{equation}
with $r=\sqrt{\left|\frac{t_{1} -\gamma}{t_{1} + \gamma}\right|}$ and $\theta \in [0, 2\pi)$. The gap-closing equations can be derived directly: $\omega = 2 n_{\beta}$ and
$R_{+}(\beta) R_{-}(\beta) =0$.

Therefore, the boundaries of the topological phases can be expressed in a format as follows:
\begin{align}
t_{1} &= \pm \sqrt{t_{2}^{2} + \gamma^{2}}, 
\nonumber\\
\omega &= 2 \left|t_{2} \pm \sqrt{\left|t_{1}^{2}- \gamma^{2}\right|} \right|.
\end{align}

As shown in figure~\ref{pd}, in which only the side $t_1>0$ have been taken, the black dotted curves are the theoretical boundaries derived through the non-Bloch band theory, and the blue solid curves are the numerical boundaries obtained by calculating the quasienergies through Floquet Hamiltonian, and the two are in good agreement.
There are four different topological phases exist in the phase diagram, which are 
(i) the topological regions where only zero mode exists $(1, 0)$, 
(ii) the topological region where both zero mode and $\pi$ mode exist $(1,1)$, 
(iii) the topological region where only $\pi$mode exists $(0,1)$, 
and (iv) the topological trivial region where neither zero mode nor $\pi$ mode exists $(0,0)$.
\begin{figure}
    \includegraphics[width=0.48\textwidth]{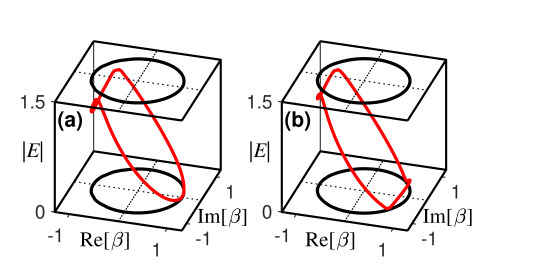}
    \caption{The quasienergies on GBZs of the gapless phases. Parameters are $\omega=3$, $\lambda=0.5$, $\gamma=0.2$,  $t_{2}=1$, with (a) $t_{1}=2.02$ and (b) $t_{1}=2.2$. Reprinted with permission from~\cite{cao2021non}, Copyright (Year) by the American Physical Society.}\label{gapless}
\end{figure}

Notably, there exists a gapless area, i.e., the gray part in figure~\ref{pd}, in the phase diagram. 
In this area, NBWNs can not be well defined. 
To elucidate the gapless phases,
the quasienergies on the GBZs of the phases are displayed in figure~\ref{gapless}.
The GBZ ($C_{\beta}$) contains two exceptional points with $E(\beta_{c_{1}}) = E(\beta_{c_{2}}) = 0$ and $\beta_{c_{1}} = \beta_{c_{2}}^{*}$. 
During the topological transition, the two EPs will merge into a single point.
\subsection{Periodically quenched non-Hermitian SSH model}\label{sec:3.3}
\begin{figure} 
		\centering
			\includegraphics[width=0.45\textwidth]{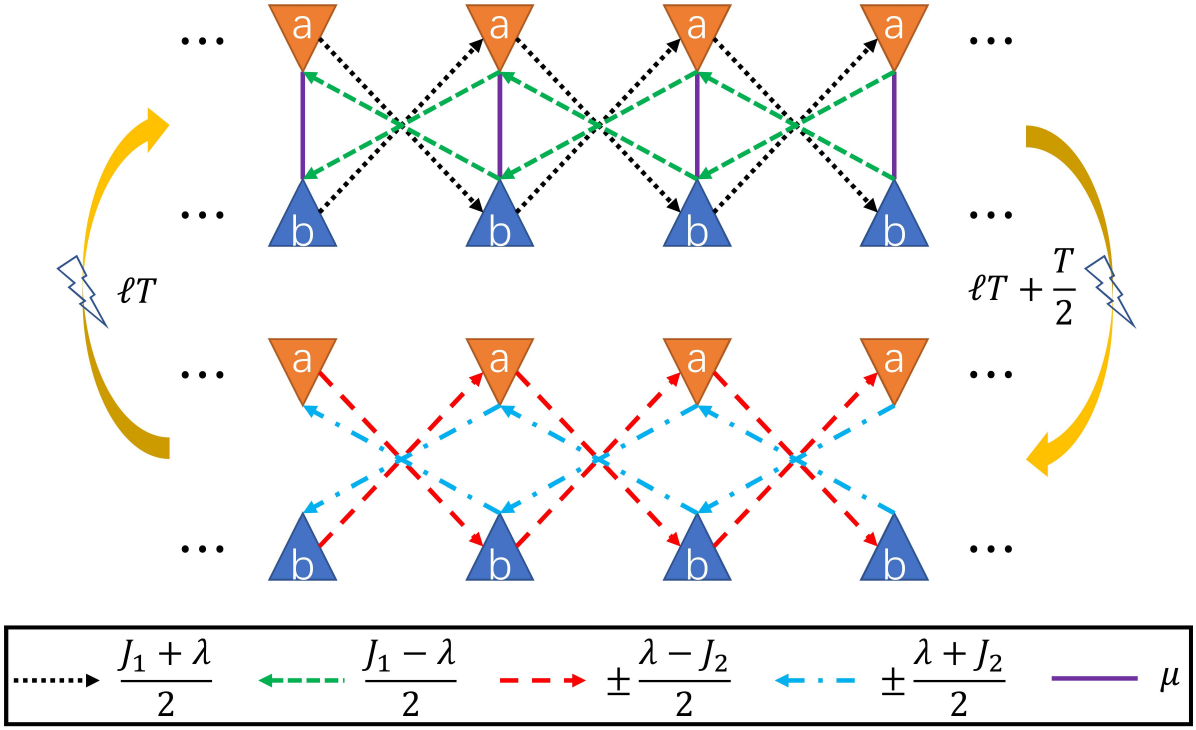} 
		\caption{A sketch
		 of the periodically quenched non-Hermitian SSH model.
			The value of intracell hopping amplitude is denoted by $\mu$. 
			During the two distinct halves of the driving period, the intercell hopping amplitudes
			are denoted as $(J_{1}\pm\lambda)/2$ and 
			$\pm(\lambda\pm J_{2})/2$, respectively.
			The lightning arrows represent quenches that are implemented during the middle and end of each driving phase.
			 Subsequently, the lattice transitions from one array. Reprinted with permission from ~\cite{zhou2021dual}, Copyright (Year) by the American Physical Society.\label{fig:quench_model}}
	\end{figure}
~~Another important type of periodically driven systems are the periodically quenched systems~\cite{prb2018zhouNon,zhou2019dynamical, zhang2020non, zhou2021dual, zhou2022q, zhou2022driving}. For these systems, the topological invariants can be defined in real space (dubbed as open-bulk topological invariants) for the periodically driven non-Hermitian systems. In the reference~\cite{zhou2021dual}, an open-boundary winding number in real space has been defined for a periodically quenched non-Hermitian SSH model to characterize the topological edge modes. Then, the generalized BBC was established for the periodically quenched non-Hermitian system.
\subsubsection{Model}

~Considering a non-Hermitian SSH model with a time-periodic quencher~\cite{zhou2021dual} sketched in figure~\ref{fig:quench_model}, the Bloch Hamiltonian can be written as following:
\begin{equation}
	H(k,t)=
	\left\{\begin{array}{l}
		(J_1\cos k+i\lambda \sin k +\mu)\sigma_x, \  t\in[0,\frac{T}{2}),\\
		(J_2\sin k+i\lambda \cos k)\sigma_y,\  t\in [\frac{T}{2},T),
	\end{array}\right.
	\label{eq:quench_hamk}
\end{equation}
with momentum $k\in [-\pi,\pi)$, driving period $T$, and Pauli matrices $\sigma_{x,y,z}$.
$\mu$ is the intracell hopping amplitude. 
$J_1$ and $J_2$ denote the strengths of intercell hopping.
The Hermiticity of the system is broken by nonreciprocal hoppings with nonzero $\lambda$.
$T=2$ is taken for simplicity.	
Then, the Floquet operator for the system of equation~(\ref{eq:quench_hamk}) can be written as:
\begin{equation}
U(k)=\exp({-ih_y\sigma_y})\exp({-ih_x\sigma_x}),
\end{equation}
with $h_y=J_2\sin k+i\lambda \cos k$ and $h_x=J_1 \cos k+i\lambda \sin k +\mu$.
The effective Hamiltonian can be defined to establish the topological characterization in the momentum space. By the same procedure of establishing  topological characterization of  periodically driven systems~\cite{zhou2018floquet, asboth2012symmetries, asboth2013bulk}, two symmetric time intervals should be introduced, where $U(k)$ hold the form of
\begin{align}
	U_{1}(k)&=e^{-i\frac{h_{x}(k)}{2}\sigma_{x}}e^{-ih_{y}(k)\sigma_{y}}e^{-i\frac{h_{x}(k)}{2}\sigma_{x}}=e^{-iH_{1}(k)},\nonumber\\
	U_{2}(k)&=e^{-i\frac{h_{y}(k)}{2}\sigma_{y}}e^{-ih_{x}(k)\sigma_{x}}e^{-i\frac{h_{y}(k)}{2}\sigma_{y}}=e^{-iH_{2}(k)}.
\end{align}
$U_{1,2}(k)$ can be obtained by applying a similarity transformation to $U(k)$. 
It ensures that they share an identical Floquet quasienergy spectrum, which is capable of being acquired through the solution of the eigenvalue equation $H_{\alpha}(k)|\psi_{\alpha}^{\pm}(k)\rangle=\pm E(k)|\psi_{\alpha}^{\pm}(k)\rangle$ for $\alpha=1,2$. 
The effective Hamiltonians can be derived by the Floquet operator and can be expressed as
\begin{equation}
		H_{\alpha}(k)=h_{\alpha x}(k)\sigma_{x}+h_{\alpha y}(k)\sigma_{y},\quad\alpha=1,2,\label{eq:Hak}
\end{equation}
where $h_{\alpha x}(k)$ and $h_{\alpha y}(k)$ are given by:
\begin{align}
	h_{1x}(k) & =\frac{E(k)\sin[h_{x}(k)]\cos[h_{y}(k)]}{\sin[E(k)]},
	\nonumber\\
	h_{1y}(k) & =\frac{E(k)\sin[h_{y}(k)]}{\sin[E(k)]},
	\nonumber\\
	h_{2x}(k) & =\frac{E(k)\sin[h_{x}(k)]}{\sin[E(k)]},
	\nonumber\\
	h_{2y}(k) & =\frac{E(k)\sin[h_{y}(k)]\cos[h_{x}(k)]}{\sin[E(k)]},\label{eq:quench_hxy}
\end{align}
with $E(k)=\arccos\{\cos[h_{x}(k)]\cos[h_{y}(k)]\}$. 
$H_{\alpha}(k)$ have the chiral symmetry $\sigma_{z}H_{\alpha}(k)\sigma_{z}=-H_{\alpha}(k)$,
the time-reversal symmetry $\sigma_{0}H_{\alpha}^{*}(k)\sigma_{0}=H_{\alpha}(-k)$ and the particle-hole symmetry $\sigma_{z} H_{\alpha}^{*}(k)\sigma_{z}=-H_{\alpha}(-k)$, where $\sigma_{0}$ is the $2\times2$ identity
matrix. 
The chiral symmetry  guarantees that the eigenenergies of $H_\alpha(k)$ arise in $(E,-E)$ pairs. Thus, the winding numbers $w_{\alpha}$ can be defined by the Bloch effective Hamiltonian~\cite{zhou2021dual}:
\begin{equation}
	w_{\alpha}=\int_{-\pi}^{\pi}\frac{dk}{2\pi}\partial_{k}\phi_{\alpha}(k), \quad\alpha=1,2, \label{eq:Wa}
\end{equation}
where $\phi_{\alpha}(k)=\arctan[h_{\alpha y}(k)/h_{\alpha x}(k)]$ is the winding angle. 
The bulk topological properties of the non-Hermitian Floquet system can be characterized by the two winding numbers defined by~\cite{zhou2021dual}:
\begin{equation}
w_{0}=\frac{w_{1}+w_{2}}{2},\qquad w_{\pi}=\frac{w_{1}-w_{2}}{2},\label{eq:W0P}
\end{equation}
which correspond to the topological zero modes and $\pi$ modes, respectively.
\begin{figure}
\begin{centering}
\includegraphics[scale=0.41]{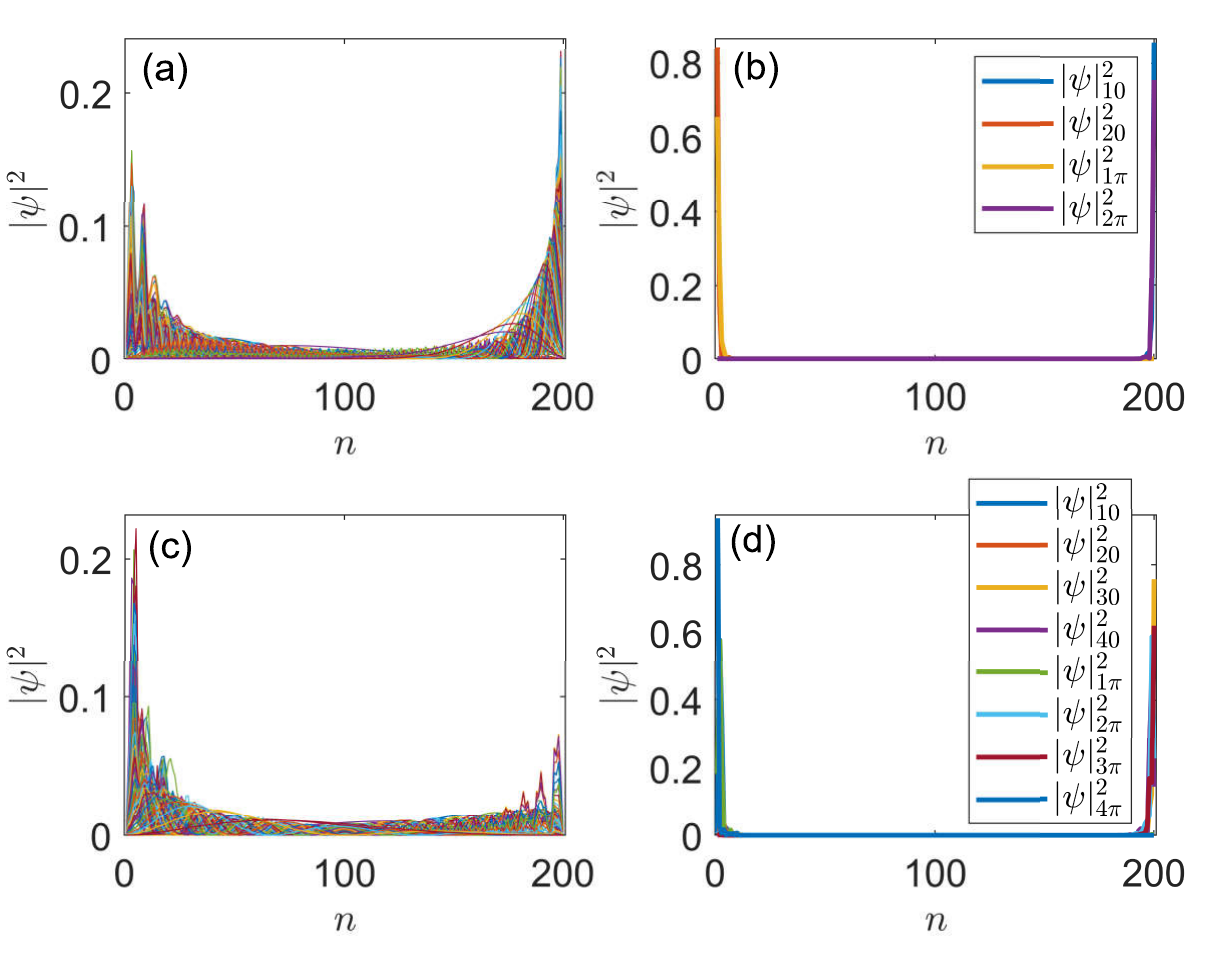}
\par\end{centering}
\caption{Profile of eigenstates of bulk [(a)(c)]and edge[(b)(d)] for the periodically quenched non-Hermitian SSH model in equation~(\ref{eq:quench_hamk}). $n$ is the index of the unit cell. 
 The number of unit cells is $200$.
Parameters are $J_2=0.5\pi, \mu=0.4\pi, \lambda=0.25$ and $J_1=\pi$ for (a)(b), $J_1=2\pi$ for (c)(d).  Reprinted with permission from~\cite{zhou2021dual}, Copyright (Year) by the American Physical Society.
\label{fig:BulkEdgeState}}
\end{figure}
	
\subsubsection{Real-space topological invariants}
The probability distributions of the bulk states are shown in figure~\ref{fig:BulkEdgeState} (a) and (c), which show that the bulk states are localized at the ends of the chain. 
Figure~\ref{fig:BulkEdgeState} (b) and (d) illustrate the corresponding distributions of topological edge modes, indicating the presence of one pair (or two pairs) of Floquet edge modes at the eigenenergies of zero and $\pi$. For the systems with the NHSE, it is useful to extend the open-bulk winding numbers from static non-Hermitian systems to periodically driven non-Hermitian systems~\cite{zhongwangPhysRevLett2018b}.

As introduced in section~\ref{sec:2.3}, the key procedure for obtaining the open-bulk winding numbers is to obtain the $Q$ matrix.
Different from the static non-Hermitian systems, the Hamiltonians of the periodically driven non-Hermitian systems contain the information of time. 
Therefore, the $Q$ matrix is defined on right (left) Floquet eigenvectors $|\psi_{\alpha n}^{\pm}\rangle$ ($\langle\tilde{\psi}_{\alpha n}^{\pm}|$) of Floquet operator in real space $U_{\alpha}$ which is the real-space representation of $U_{\alpha}(k)$.
The left and right Floquet eigenvectors are defined by the eigenvalue equations $U_{\alpha}|\psi_{\alpha n}^{\pm}\rangle=e^{-i(\pm E_{n})}|\psi_{\alpha n}^{\pm}\rangle$ and $\langle\tilde{\psi}_{\alpha n}^{\pm}|U_{\alpha}=\langle\tilde{\psi}_{\alpha n}^{\pm}|e^{-i(\pm E_{n})}$, respectively, where $\pm E_n$ are the eigenenergies. 
Then, the $Q$ matrix can be defined as:
\begin{equation}
	Q_{\alpha}=\sum_{n}(|\psi_{\alpha n}^{+}\rangle\langle\tilde{\psi}_{\alpha n}^{+}|-|\psi_{\alpha n}^{-}\rangle\langle\tilde{\psi}_{\alpha n}^{-}|). 
\end{equation}
With a well-defined $ Q_{\alpha}$, the open-bulk winding numbers for periodically driven systems~\cite{zhou2021dual} can be defined as
\begin{equation}
	 W_{\alpha}=\frac{1}{L_{\rm B}}{\rm Tr}_{\rm B}({\cal C} Q_{\alpha}[ Q_{\alpha},{\cal N}]).\label{eq:NUa}
\end{equation}
Here, ${\cal C}=\mathbb{I}_{N\times N}\otimes\sigma_{z}$ is the operator of the chiral symmetry , $\mathbb{I}_{N\times N}$ is an $N\times N$ identity matrix and $N$ denotes the total number of unit cells. 
$L_{\rm B}$ and ${\rm Tr}_{\rm B}$ have the same physical meanings as the static open-bulk winding numbers.
It means that the trace  ${\rm Tr}_{\rm B}$ is taken over the bulk area, which contains $L_{\rm B}$ lattice sites after the system has been divided into a bulk region and two edge regions along the left and right boundaries.
In the periodically quenched non-Hermitian SSH model, there exist two different types of topological states, i.e., zero modes and $\pi$ modes, which need two  different types of open-bulk winding numbers $W_{0}, W_{\pi}$~\cite{zhou2021dual} defined as:
\begin{equation}
	W_{0}=\frac{W_{1}+W_{2}}{2},\qquad W_{\pi}=\frac{W_{1}-W_{2}}{2}.\label{eq:NU0P}
\end{equation}
Here, $W_{0}$ and $W_{\pi}$ are the open-bulk winding numbers, which can characterize the topological zero modes and $\pi$ modes, respectively.
Then the generalized BBC for periodic quenched non-Hermitian SSH model is established with the relation $(n_{0},n_{\pi})=2(|W_{0}|,|W_{\pi}|)$, in which $n_{0}$ ($n_{\pi}$) is the number of the topological zero ($\pi$) modes.
 \begin{figure}
    \begin{centering}
        \includegraphics[width=0.45\textwidth]{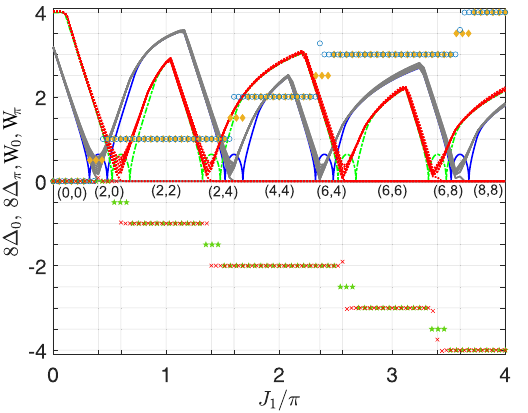}
        \par\end{centering}
\caption{The gap functions $\Delta_{0}$ (gray/blue solid curves) and $\Delta_{\pi}$ (red/green dotted curves) versus $J_{1}$ for the model in equation~\ref{eq:quench_hamk} under OBC/PBC. The open-bulk winding numbers $W_{0}$ and $W_{\pi}$ are labled by circles and crosses curves. The PBC winding numbers $w_0$ and $w_\pi$ are labeled by diamonds and pentagrams curves.  
For clarity, only the twenty smallest gap functions $(\Delta_{0},\Delta_{\pi})$ are shown.
The numbers of the topological zero modes and $\pi$ modes are denoted by ($n_0, n_\pi$).
Other parameters are $\mu=0.4\pi$, $J_2=0.5\pi$ and $\lambda=0.25$. Reprinted with permission from~\cite{zhou2021dual}, Copyright (Year) by the American Physical Society. \label{fig:quench_bbc}}
\end{figure}

Figure~\ref{fig:quench_bbc} depicts the comparison of the eigenenergies around the two gaps $(\Delta_{0},\Delta_{\pi})$ and the winding numbers of the periodically quenched non-Hermitian SSH model under the OBC and the PBC.
The gap functions were defined to clearly reveal the gaps as $\Delta_{0}=|E|/\pi$ for zero gap and $\Delta_{\pi}=\sqrt{(|\pi-{\rm Re}E|)^{2}+({\rm Im}E)^{2}}/\pi$ for $\pi$ gap. 
The gap of the spectra closes with $\Delta_{0}=0$ ($\Delta_{\pi}=0$) at zero gap ($\pi$ gap), where there can exist a phase transition. 
Figure~\ref{fig:quench_bbc} shows $(\Delta_{0},\Delta_{\pi})$ varies with hopping amplitude $J_{1}$  of a periodically quenched non-Hermitian SSH model with $L=400$ sites under both the PBC and the OBC.
The PBC and the OBC energy spectra are distinct at the gap closing points. 
For example, the spectrum under the OBC indicates a phase transition at $J_1=0.4\pi$, followed by the emergence of a pair of topological zero edge modes, while the PBC energy spectrum implies two sequential transitions at $J_{1}<0.4\pi$ and $J_{1}>0.4\pi$. 
The significant disparity in the Floquet spectrum between systems under the PBC and OBC implies the existence of NHSE and the failure of the Bloch BBC.

The open-bulk winding numbers $(W_{0/\pi})$, which can be calculated from equations~(\ref{eq:NUa})-(\ref{eq:NU0P}), are shown in figure~\ref{fig:quench_bbc}.
The values of $W_{0/\pi}$ are integers for Floquet non-Hermitian topologically nontrivial phases.
These values suffer quantized jumps at a topological phase transition with gap closing under the OBC. 
$W_0$ and $W_\pi$ accurately precisely quantify the amount of zero modes and $\pi$ modes, establishing the generalized BBC of the periodically driven non-Hermitian systems despite the presence of the NHSE. 
\begin{figure*} 
	\begin{centering}
		\includegraphics[width=0.8\textwidth]{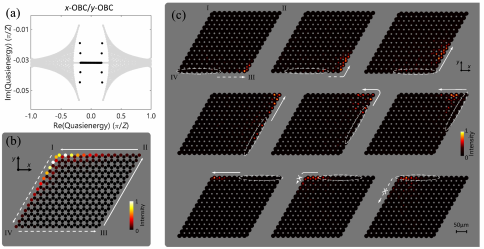}
		\par
	\end{centering}
	\caption{(a) Quasienergy spectrum of the system under OBC in both directions. The black dots denoted the skin-topological modes.
		(b) The distribution of the skin-topological modes.
		(c) Experimental observation of the Floquet skin-topological effect. The output light is distributed at the end facet of the sample after a 10 cm long propagation by moving the input tilted Gaussian beam along the outer perimeter. Reproduced from~\cite{sun2023photonic}. CC BY 4.0..
		\label{fig:fste}}
\end{figure*}
\subsection{Higher-dimensional Floquet non-Hermitian systems}\label{sec:3.4}
Higher-dimensional non-Hermitian  systems offer diverse systems for the realization of the rich novel non-Hermitian phenomena, topological states,  and their mutual hybridization. 
The NHSE not only can act on the bulk states but also influence the topological edge states.
The interaction between topological edge modes and NHSE in two/three- dimensional systems can lead to the accumulation of topological edge modes at lower-dimensional boundaries. 
This phenomenon is referred to as the hybrid skin-topological effect~\cite{zhu2022hybrid,lee2019hybrid, fu2021non, kawabata2020higher, gao2021non, li2022gain,  ghosh2022non}. 
An intriguing aspect of the skin-topological effect is that the NHSE can act only on the topological edge mode, while the bulk states remain extending.
Apart from the NHSE, the higher-order topological phases~\cite{EdvardssonPhysRevB2019, gao2021non,ghosh2022non,liu2019second, zhu2019second, luo2019higher, EzawaPhysRevB2019} also enable the emergence of topological edge states in lower dimensions: A nth-order topological insulator or superconductor with d spatial dimension can have (d-n)-dimensional topological gapless states.
The NHSE in a higher-dimensional non-Hermitian system can also depend on the system's geometry under OBC, the phenomenon of which is referred to as the geometry-dependent NHSE~\cite{zhang2022universal,fang2022geometry}.

Very recently, the general non-Bloch band theory in higher-dimensional non-Hermitian systems was established~\cite{wang2022amoeba,zhang2023edge}.
With the help of the non-Bloch band theory, the generalized BBC can be established for a higher-dimensional system with the NHSE. 
Periodic driving provides intriguing possibilities for enriching novel phenomena in higher-dimensional systems with high tunability.
The interplay between the periodic driving and the topology in higher-dimensional non-Hermitian systems has been only partially revealed, and further extensive research is still awaited. 
Here we present the photonic Floquet skin-topological effect~\cite{sun2023photonic} and the Floquet second-order topological insulator as illustrations.

\subsubsection{Floquet hybrid skin-topological effect}
Recently, the Floquet skin-topological effect was realized by Yang's group in a two-dimensional (2D) photonic Floquet non-Hermitian topological insulator~\cite{sun2023photonic}.
Initially, a 1D optical array was constructed, comprising helical waveguides that incorporate two sublattices, namely A and B, with significant loss in sublattice B.
The lD optical array is experimentally realized by the method of femtosecond laser writing. The optical loss in sublattice B occurs due to the intentional periodic insertion of breaks into the waveguides.
The Floquet NHSE, resulting from periodic driving and loss, can be detected in the 1D optical array. 
This observation serves as a fundamental basis for further investigation into the interaction between the NHSE and photonic topological edge states.
The 2D photonic Floquet non-Hermitain topological insulator is constructed by stacking this 1D Floquet optical array.
The quasienergy spectrum of the 2D photonic NHFTI with nontrivial Chern topology under OBC in both directions is shown in figure~\ref{fig:fste}(a), in which the energies of bulk and chiral edge states are denoted by gray and black dots, respectively. 
The corresponding distribution of chiral edge modes is depicted in figure~\ref{fig:fste}(b). 
The accumulation of these edge modes occurs at a corner of the lattice as a result of the hybridization of NHSE and topology. 
This phenomenon was referred to as the Floquet skin-topological effect.
The experimental observation was predicated on a honeycomb lattice composed of helical waveguides.
A Gaussian beam, indicated by the white ellipse dashed curves, is emitted into the outer boundary of the sample to stimulate the topological edge modes, which was shown in figure~\ref{fig:fste}(c).
As the injected beam traversed the outer perimeter (IV$\rightarrow$III$\rightarrow$II), the light propagated along the edge without experiencing backscattering, even in the presence of a sharp corner.
In the case where the beam is moved along II$\rightarrow$I, the injected light travels along the upper edge and remains confined at corner I without entering the bulk. 
Through the topological phase transition, the topological switch of Floquet skin-topological effect can be observed~\cite{sun2023photonic}.

\subsubsection{Floquet non-Hermitian second-order topological insulator}
\begin{figure*} 
	\begin{centering}
		\includegraphics[width=0.85\textwidth]{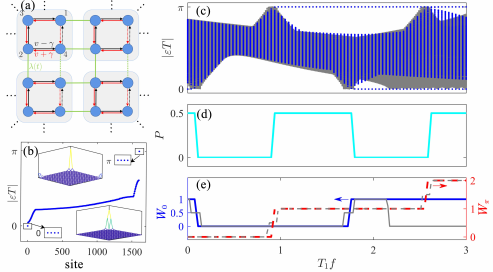}
		\par
	\end{centering}
	\caption{(a)The sketch of Floquet non-Hermitian second-order topological insulator.The red and black arrows denote the intracell hopping. 
	The green lines represent the intercell hopping.(b) The quasienergy spectrum of a 20$\times$20 square lattice under OBC in both directions. The insets depict the probability distributions of topological zero modes (the upper inset) and $\pi$ modes (the lower inset), respectively.
	Parameters in (b) are set as $T_1=T_2=2.2 f^{-1}$ $v=0.8 f$, $\gamma=0.4 f$, $q_1=1$, and $q_2=0$.
		(c) The quasienergy spectrum with varied $T_1$ of the FNHSOTI under OBC (marked in blue) and PBC (marked in gray), respectively.
		(d) The quadruple moment varied with $T_1$.
		(e) The blue and red lines denote the winding numbers at two gaps varied with $T_1$ based on GBZ, respectively. 
		The gray line represents the BZ solution.
		Phase diagram based on the winding numbers $W_1$ and $W_2$.
		Parameters in (c)--(e) are $T_2=0.3f^{-1}$, $v=1.2f$, $\gamma=0.2f$, $N=30$, and $q_1=-q_2=1.5f$. Reprinted with permission from~\cite{wu2021floquet_SOTI}, Copyright (Year) by the American Physical Society.
		\label{fig:fsoti}}
\end{figure*}

Here we introduce a 2D Floquet non-Hermitian second-order topological insulator (FNHSOTI), in which the Bloch BBC is broken down~\cite{wu2021floquet_SOTI}.
A sketch of this model is depicted in figure~\ref{fig:fsoti}(a), and the Bloch Hamiltonian takes the form of
\begin{align}
	&H_{2 \mathrm{D}}(k_x,k_y,t)
	\nonumber\\
	&= [v+\lambda(t) \cos k_x] \tau_x \sigma_0-[\lambda(t) \sin k_x+i \gamma] \tau_y \sigma_z +
	\nonumber\\
	&\quad  [v+\lambda(t) \cos k_y] \tau_y \sigma_y+[\lambda(t) \sin k_y+i \gamma]\tau_y \sigma_x.\label{eq_fsoti_model}
\end{align}
Here $\tau_i$ and $\sigma_i$ are Pauli matrix, and $v\pm\gamma$ denotes the strength of nonreciprocal intracell hopping.
The strength of intercell hopping amplitude  is periodically quenched with the following form
\begin{align}
	\lambda(t)=\left\{\begin{array}{l}
		\lambda_1=q_1 f, t \in\left[m T, m T+T_1\right)  ,
		\\
		\lambda_2=q_2 f, t \in\left[m T+T_1,(m+1) T\right) .
	\end{array}\right.
\end{align}

The generalized BBC of the periodically driven non-Hermitian systems can be constructed by the non-Bloch band theory based on the GBZ.
The non-Bloch Hamiltonian $H_{\mathrm{2D}}(\tilde{k}_x,\tilde{k}_y,t)$ can be obtained by replacing Bloch factors $e^{ik_{x/y}}$ by the non-Bloch factors $e^{i\tilde{k}_{x/y}}$ where $\tilde{k}_{x/y}=k_{x/y}- i\log(r_{x/y})$ ($k_{x/y}$ and $r_{x/y}$ are real).
Here $r_{x/y}=\sqrt{(v-r)/(v+r)}$ can be obtained through the method of similarity transform~\cite{wu2021floquet_SOTI, zhongwangPhysRevLett2018a}.
Based on the Floquet theorem, the Floquet operator takes the form of 
\begin{equation}
	U_{T}(\tilde{k}_x,\tilde{k}_y)=e^{-iH_2(\tilde{k}_x,\tilde{k}_y)T_2}e^{-iH_1(\tilde{k}_x,\tilde{k}_y)T_1},
\end{equation}
where $H_1(\tilde{k}_x,\tilde{k}_y)=H_{\mathrm{2D}}(\tilde{k}_x,\tilde{k}_y,t)|_{\lambda(t)=q_1f}$ and $H_2(\tilde{k}_x,\tilde{k}_y)=H_{\mathrm{2D}}(\tilde{k}_x,\tilde{k}_y,t)|_{\lambda(t)=q_2f}$.
The effective Hamiltonian can be derived from the Floquet operator via $H_{\mathrm{eff}}(\tilde{k}_x,\tilde{k}_y)=i\log U_T(\tilde{k}_x,\tilde{k}_y)/T$ and satisfies the mirror-rotation symmetry $M_{xy}H_{\mathrm{eff}}(\tilde{k}_x,\tilde{k}_y)M_{xy}^{-1}=H_{\mathrm{eff}}(\tilde{k}_y,\tilde{k}_x)$ with $M_{xy}=[(\tau_0-\tau_z)\sigma_x-(\tau_0+\tau_z)\sigma_z]/2$.
The quasienergy spectrum of the system is shown in figure~\ref{fig:fsoti}(b), in which four topological zero modes and four topological $\pi$ modes can emerge in the two types of gaps. The probability distributions of topological zero modes and $\pi$ modes are depicted in the inset of figure~\ref{fig:fsoti}(b), respectively.
Topological zero modes and $\pi $ modes are sharply localized at the corners of the system.

Although the instantaneous Hamiltonian satisfied chiral symmetry $\mathcal{C}$ with $\mathcal{C}H(\tilde{k}_x,\tilde{k}_y,t)\mathcal{C}^{-1}=-H(\tilde{k}_x,\tilde{k}_y,t)$,
the effective Hamiltonian  $H_{\mathrm{eff}}$ does not have the chiral symmetry because of
\begin{equation}
	 [H_1(\tilde{k}_x,\tilde{k}_y), H_2(\tilde{k}_x,\tilde{k}_y)] \neq 0.
\end{equation}
The chiral symmetry can be restored by converting $U_{T}(\tilde{k}_x,\tilde{k}_y)$ into $U_1(\tilde{k}_x,\tilde{k}_y)$ and $U_2(\tilde{k}_x,\tilde{k}_y)$ through the similarity transformation $G_j(\tilde{k}_x,\tilde{k}_y)=\exp[i(-1)^j H_j(\tilde{k}_x,\tilde{k}_y)T_j /2]$, $j=1$, $2$. The corresponding effective Hamiltonians are given by $H_{\mathrm{eff},j}(\tilde{k}_x,\tilde{k}_y)=i\log U_j(\tilde{k}_x,\tilde{k}_y)/T$.
Due to the mirror-rotation symmetry, the topological properties of FNHSOTIs can be characterized by the winding numbers $W_j$ based on the effective Hamiltonian along the high-symmetric line $H_{\mathrm{eff},j}(\tilde{k},\tilde{k})$ through
\begin{equation}
	W_{j}=\frac{1}{2\pi i}\int_{C_\beta} \mathrm d  \tilde k\frac{\mathrm d }{\mathrm d  \tilde k} \log \det H_{\mathrm{eff,j}}(\tilde k,\tilde k),\quad j=1,2.
\end{equation}
The winding numbers correspond to topological zero modes and $\pi$ modes can be constructed through~\cite{wu2021floquet_SOTI}
\begin{equation}
	W_0= 2\abs{W_1+W_2}, \quad W_{\pi}= 2\abs{W_1-W_2}.
\end{equation}

The quadrupole moment, which is widely used in characterizing static second-order topological insulators, should be extended by the biothogonal basis in a non-Hermitian system. For a non-Hermitian system whose Hamiltonian is $H$, the left and right eigenvector are given by $H\left|\psi_n^{\mathrm{R}}\right\rangle=E_n\left|\psi_n^{\mathrm{R}}\right\rangle$ and $H^{\dagger}\left|\psi_n^{\mathrm{L}}\right\rangle=E_n^*\left|\psi_n^{\mathrm{L}}\right\rangle$. The quadrupole moment for a non-Hermitian square lattice under OBC in both directions~\cite{wu2021floquet_SOTI} can be defined as
\begin{equation}
	P=\left[\frac{\operatorname{Im} \ln \operatorname{det} \mathcal{U}}{2 \pi}-\sum_{\mathbf{n}, i ; \mathbf{m}, j} \frac{X_{\mathbf{n}, i ; \mathbf{m}, j}}{2 L_x L_y}\right] \bmod 1.
\end{equation}
Here the elements of $\mathcal{U}$ hold the form of $\mathcal{U}_{\alpha \beta} \equiv \bra{\psi_\alpha^{\mathrm{L}}} e^{i 2 \pi X /\left(L_x L_y\right)} \ket{\psi_\beta^{\mathrm{R}}}$ and $X_{\mathbf{n}, i; \mathbf{m}, j}=n_x n_y \delta_{\mathbf{n m}} \delta_{i j}$ denotes the coordinate, in which $i=1,\cdots,4$ represent the sublattices and $L_{x/y}$ are the length of system in $x/y$ directions.

 In the FNHSOTI described by equation~\ref{eq_fsoti_model}, the quadrupolar moment and the winding numbers have the following relation~\cite{wu2021floquet_SOTI}:
\begin{equation}
	P=\frac{1}{2}\left(\left|W_0\right|+\left|W_{\pi}\right|\right) \bmod 2.
\end{equation}
The numerical results of the quasienergy spectrum of varied $T_1$ are shown in figure~\ref{fig:fsoti}(c). The corresponding quadruple moment (cyan line) and the winding numbers based on GBZ (blue and red lines) and BZ (gray lines) are shown in figures~\ref{fig:fsoti}(d) and (e), respectively.
When comparing the quadruple moment to the winding numbers based on GBZ, it is evident that the latter is more fundamental. 
An example that demonstrates this is the scenario when $T_1f=2$.
In this case, the quadruple moment ($P$) is topologically trivial, while winding numbers based on GBZ ($W_{0/\pi}$) are nonzero and accurately predict the topological edge modes.
The winding numbers based on the BZ fail to predict the emergence of the topological modes in FNHSOTI, indicating the breaking down of conventional BBC based on Bloch theory.
The winding numbers based on GBZ provided an accurate prediction of the topological zero modes and $\pi$ modes and established the generalized BBC for the FNHSOTIs.

\subsection{Experimental realizations}\label{sec:3.5}
The intriguing non-Hermitian or Floquet phenomena have been extensively studied in several physical platforms, including photonics/optics~\cite{ruter2010observation,midya2018non,science2020sebastian, XiaoLeiNaturePhysics2020, prr2021wuFloquet,weidemann2022topological}, acoustic~\cite{zhang2021acoustic,prl2023zhang_Braid,zhang2021observation,prb2022gaoAnomalous}, electric circuits~\cite{prl2019hofmannChiral, HofmannPRR2020,pra2020liuGain,research2021liuNon}, cold atoms~\cite{prl2022qianDynamic}, mechanical metamaterials~\cite{pnac2020ananyaObservation,wang2022non}, and nitrogen-vacancy centers~\cite{prl2021zhangObservation,yu2022experimental,wu2023observation}.
In this section, we provide a concise overview of the experimental realizations and progress in non-Hermitian and periodically driven systems,  particularly in the fields of photonics/optics, acoustics, and electric circuits.
\subsubsection{Photonics or optics}
The experimental investigation of non-Hermitian systems employing optics is one of the key areas of interest in non-Hermitian physics. Initially, Xue's group experimentally observed the NHSE and confirmed the generalized BBC in discrete-time non-unitary quantum-walk dynamics of single photons~\cite{XiaoLeiNaturePhysics2020}.
A typical method to introduce non-Hermiticity in optical systems is gain and loss.
In coupled-resonator optical waveguides ~\cite{hafezi2011robust}, gain and loss are usually achieved by pump-controlled or gate-controlled optical absorption~\cite{midya2018non}.
In photonic crystals~\cite{butt2021recent}, the loss is generated by materials having a complex permittivity or complex refractive index~\cite{prl2009guo_observation}, and gain is introduced through two-wave mixing employing the material's photorefractive nonlinearity~\cite{ruter2010observation}. 

Besides, many studies focus on the application of non-Hermitian optical systems, which can be realized based on the interaction between nonlinear optics and non-Hermitian systems. 
For instance, the transition between PT symmetry and non-PT symmetry regimes and the maneuvering of topological zero modes can be realized by adjusting the nonlinearity~\cite{xia2021nonlinear}. Adjustment of nonlinear optics also enables high-speed manipulation of multiple topological phases~\cite{dai2023non}.
Based on the NHSE, a light funnel has been developed, which is an optical device that has an appealing impact on light field research~\cite{science2020sebastian} and can be evaluated with two fiber loops~\cite{prl2023adiyatullinTopo}.

The Floquet photonic crystals can also be achieved by means of spiral dielectric columns~\cite{nsture2013rechtsman}.
The combination of periodical driving and non-Hermiticity in optical systems can lead to a variety of fascinating phenomena.
In periodically curved waveguide arrays, the introduction of gain and loss can reopen the gap of the quasienergy bands and induce Floquet $\pi$ modes~\cite{prr2021wuFloquet}.
An investigation was conducted on Floquet dissipative quasicrystals using photonic quantum walks in coupled fiber loops and has revealed an intriguing occurrence of topological triple phase transition~\cite{weidemann2022topological}.
As mentioned before, the optical Floquet skin-topological effect was observed by Yang's group~\cite{sun2023photonic}.

\subsubsection{Acoustics}
In acoustics, the non-reciprocity can be realized by introducing directional amplifiers between resonators~\cite{zhang2021acoustic,prl2023zhang_Braid}. Applying the non-reciprocity to the coupling,  the  NHSE has been observed~\cite{zhang2021acoustic}.
Non-reciprocity can also be realized in coupled resonator acoustic waveguides~\cite{zhang2021observation}. 
In particular, the larger and smaller ring-shaped waveguides serve as site whisper-gallery acoustic resonators and couplers between site resonators, respectively. 
Such waveguides support clockwise and anti-clockwise acoustic whispering-gallery modes at the same time. 
By adding biased loss in the coupled resonators along a direction, one of the modes experiences loss while the other does not when they travel along this direction, and thus the anisotropic coupling is realized. By designing a 2D acoustic higher-order topological insulator composed of coupled resonator acoustic waveguides, the spin-polarized higher-order non-Hermitian skin effect has been observed~\cite{zhang2021observation}.

Floquet topological insulators can be realized in acoustic systems through distinct methods.
Through the temporal modulation, a method to realize the Floquet topological insulator is proposed in a hexagonal-lattice acoustic crystal constructed by coupled acoustic trimers~\cite {fleury2016floquet}.
Each trimer can be perceived as a resonant acoustic metamolecule consisting of three acoustic chambers interconnected by cylindrical waveguides.
The cavity resonance is operated below the first dipolar cavity resonance, and thus the trimer is equivalent to a L-C resonating loop.
An effective spin, which breaks the time-reversal symmetry, can be carried to the trimer by modulating the acoustic capacitance of each cavity in time.
Another method is to map the temporal modulation to wave evolution in space dimension, through which the acoustic Thouless pumping~\cite{long2019floquet}, the acoustic analog of Chern insulators~\cite{prl2019pengChirality}, acoustic $\pi/2$ modes ~\cite{prl2022chengObervation}, and acoustic Floquet higher-order topology ~\cite{zhu2022time} have been experimentally realized. 
The third type is proposed to realize an optical analog of the quantum spin Hall effect by employing a 2D coupled resonator waveguide~\cite{nsture2013rechtsman}. 
The lattice is specifically designed to be periodic, without requiring an external drive to realize temporal modulation, and is called the anomalous Floquet topological insulator.
Similar structures have been adopted to realize anomalous Floquet topological insulators~\cite{peng2016experimental}.
Recently, non-Hermitian topology has also been introduced to Floquet physics. 
By designing a lattice of coupled ring resonators with uniform loss, an anomalous non-Hermitian skin effect has been demonstrated~\cite{prb2022gaoAnomalous}. 
The non-Hermitian skin modes arise from the nonreciprocal coupling induced by an additional ring resonator in each unit cell. 
Furthermore, due to the strong coupling provided by the coupled ring resonators, the designed lattice is a periodically driven system, which largely broadens the working bandwidth of the non-Hermitian skin modes. 

\subsubsection{Electric circuits}
Electric circuits are one of the most fascinating topics in topological physics research these days, as they provide a classical platform for simulating tight-binding models. 
In electric circuits, resistance is regarded as a non-Hermitian element that contributes to dissipation, whereas passive components like capacitors and inductors function as Hermitian elements~\cite{HofmannPRR2020}. 
Adding resistance to the lattice model can be thought of as adding a bidirectional complex transition between the lattice points~\cite{EzawaPhysRevB2019}. 
By constructing the negative resistance module, the gain can be introduced to the electric circuits~\cite{Schindler_2012}. 
Based on the negative resistances, the PT symmetry transition can be simulated in electric circuits~\cite{pra2011schindler, pra2012ramezani}. 
The simulation of non-Hermitian topological states caused solely through gain and loss has also been realized by many researchers. 
For example, by varying positive and negative resistance, different topological phases can be observed on a 1D non-Hermitian model~\cite{pra2020liuGain}.  
Electric circuits can also be used as operational amplifiers to simulate the non-Hermitian tight-binding model with non-reciprocal hopping. The most widely used of them is the negative impedance converter with current inversions (INIC), which can produce equal and opposite currents. INICs were employed in electric circuits to break the time-reversal symmetry~\cite{prl2019hofmannChiral}. 
The NHSE and the failure of conventional BBC were identified when they applied INICs to the non-Hermitian SSH model~\cite{HelbigNaturePhysics2020}. 
Voltage followers are used to realize non-reciprocal hopping, and paralleling the voltage follower with the passive element is a comparatively reliable method for unidirectional hopping models. 
Zhang's group demonstrated experimentally the existence of non-Hermitian peeling effect in a non-reciprocal coupled SSH model using a voltage follower~\cite{research2021liuNon}. So far, a variety of non-Hermitian phenomena have been observed on electric circuits, such as the evolution of non-Bloch waves~\cite{prb2023wuEvidencing}, higher-order non-Hermitian skin effects~\cite{nature2021zouobservation, shang2022experimental}, and many-body non-Hermitian skin effects~\cite{prb2022zhangObservation}. 

The electric circuits provide a platform to investigate the periodically driven non-Hermitian systems. For example, the periodically quenched systems can be realized by adding switches in electric circuits~\cite{chitsazi2017experimental,li2018pt}. 
Through this technique, conversion can be achieved in a single cycle with or without energy exchange. 
Kottos’s group have realized a periodically driven PT-symmetric system~\cite{leon2018observation, bahmani2023anomalous}, which consists of a combination of two coupled LC resonators with balanced gain and loss. The capacitance coupling of the two resonators is driven by a network of varactor diode. The PT symmetry phase can be modulated by the amplitude and frequency of the driving~\cite{leon2018observation}. 
Therefore, electric circuits provide a tunable classical platform for non-Hermitian topological phenomena.
\section{Conclusion and outlook}\label{sec:4}
In this review, we have provided an overview of the generalized BBC in periodically driven non-Hermitian systems by taking the periodically driven non-Hermitian SSH model as a paradigm. 
Specifically, two typical periodically driven non-Hermitian systems have been introduced, i.e., the harmonically driven non-Hermitian SSH model and the periodically quenched non-Hermitian SSH model. In a harmonically driven non-Hermitian SSH model with the NHSE, the NBWNs based on the GBZ have been established to characterize the two distinct types of topological edge modes. In addition, the non-Bloch band invariants based on the non-Bloch effective Hamiltonian have been developed as a comparison. The non-Bloch band theory, in which the GBZ plays an essential role, provides a useful tool to investigate the topological states of the periodically driven non-Hermitain system. In some non-Hermitan systems, it is difficult to obtain the GBZ, in which case the open-bulk topological invariants can be defined in real space to predict the topological edge states. A periodically quenched non-Hermitian SSH model with the NHSE has been introduced to studying the open-bulk winding numbers as the topological characterization. 
We also have viewed the topological properties of higher-dimensional non-Hermitian systems with periodic driving by a focus on introducing the Floquet hybrid NHSE and the Floquet second-order topological insulators.
Additionally, we have surveyed the experimental realizations in optics, acoustics, and electric circuits. 

In the near future, there will be further developments in the periodically driven non-Hermitian systems, since many issues still require more theoretical and experimental efforts. 
For instance, the topological properties of the periodically driven non-Hermitian superconductors with particle-hole symmetry. The non-Hermitian topological superconductors have exhibited many interesting phenomena. The NHSE corresponds to the $Z_2$ skin effect, and the phase transition occurs at the Bloch points~\cite{li2022universal}. The interplay between the periodic driving and the particle-hole symmetry may display rich interesting phenomena and this remains to be investigated so far. The periodic driving will enrich the topological properties of the Majorana corner modes and the topological edge states in non-Hermitian higher-order superconductors. Furthermore, the energy spectrum forming the Wannier-Stark ladder will induce the Wannier-Stark localization of the eigenstates ~\cite{PRBWS1987}. The periodic driving can induce the Wannier-Stark localization along the Floquet direction in the Hermitian~\cite{oka2019floquet} and non-Hermitian systems~\cite{WangJPCM2022NH}. The competition between the non-Hermitian skin effect and Wannier-Stark localization in non-Hermitian systems has generated many interesting phenomena~\cite{WangPRA2022ws, WangJPCM2022NH, ZengPRB2023ws}. The interplay between the non-Hermitian skin effect and the Wannier-Stark localization can enrich the generalized bulk-boundary correspondence for the periodically driven non-Hermitian topological phases and seems to be a promising future avenue. 
We hope this pedagogical article will motivate further research on the Floquet non-Hermitian topological physics.
\section*{Data availability statement}
All data that support the findings of this study are included within the article (and any supplementary files). We have provided the code for calculating the GBZ of the periodically driven SSH model in the supplementary file.
\section*{Acknowledgements}
We thank Prof. Z. Wang for fruitful discussions. This work was supported by the Natural Science Foundation of Jiangsu Province (Grant No. BK20231320) and National Natural Science Foundation of China (Grant No. 12174157).

\bibliographystyle{iopart-num-mod.bst} 
\bibliography{ref.bib} 
\onecolumn
\appendix
\section*{Matlab code for calculating the GBZ of periodically driven SSH model}
\begin{lstlisting}
% The code is cheked with Malab 2021b
clear; clc
syms B
L=40;% size of the hamiltonian
t1=0.75;
gamma=0.2;% strength of non-Hermiticity
t2=1.0;t3=0; 
lambda=0.5;% strength of periodic driving
w=3; % angular frequency
T=2*pi/w; 
floquet_number=3;

Energy_midband = floquetHammltonian(t1,t2,t3,gamma,lambda,w,L,floquet_number);
% quasienergy in OBC over one period
Cbeta = GBZ(t1,t2,t3,gamma,lambda,w,Energy_midband,B);
figure()
subplot(2,1,1)
scatter(real(Energy_midband),imag(Energy_midband),'.');
title(gca,'Energy spectrum');
xlabel('Re(\epsilon)');ylabel('Im(\epsilon)');
subplot(2,1,2)
scatter(real(Cbeta),imag(Cbeta),'.')
title(gca,'GBZ');
xlabel('Re(\beta)');ylabel('Im(\beta)');		
		
function Energy_midband=floquetHammltonian(t1,t2,t3,gamma,lambda,w,L,floquet_number)
NX=L;NY=1;NXY=NX*NY;
delta=0.01;
hamiltonian(2*floquet_number*NXY,2*floquet_number*NXY)=zeros;
for ii=0:floquet_number-2
	for ix=1:NX
		i=ix+ii*2*NXY;
		hamiltonian(2*NXY+i,NXY+i)=0.5d0*lambda;
		hamiltonian(3*NXY+i,i)=0.5d0*lambda;
		hamiltonian(i,3*NXY+i)=0.5d0*lambda;
		hamiltonian(NXY+i,2*NXY+i)=0.5d0*lambda;
	end
end
for ii=0:floquet_number-1
	for ix=1:NX
		i=ix+ii*2*NXY;
		hamiltonian(i,i)=w*((floquet_number-1)*0.5d0-ii);
		hamiltonian(NXY+i,NXY+i)=w*((floquet_number-1)*0.5d0-ii);
		hamiltonian(i,NXY+i)=t1+gamma;
		hamiltonian(NXY+i,i)=t1-gamma;
	end
	for ix=1:NX-1
		i=ix+ii*2*NXY;
		hamiltonian(NXY+i,i+1)=complex(t2,0.0d0);
		hamiltonian(i,NXY+i+1)=complex(t3,0.0d0);
	end
	for ix=2:NX
		i=ix+ii*2*NXY;
		hamiltonian(NXY+i,i-1)=complex(t3,0.0d0);
		hamiltonian(i,NXY+i-1)=complex(t2,0.0d0);
	end
end

E(1,:)=eig(hamiltonian);
kk=0;
for k=1:size(E,2)
	if real(E(1,k))<w/2+delta&&real(E(1,k))>-w/2-delta
		kk=kk+1;
		Energy_midband(1,kk)=E(1,k);
	elseif floquet_number==1
		kk=kk+1;
		Energy_midband(1,kk)=E(1,k);
	end
end
end
		
% compute the GBZ
function Cbeta=GBZ(t1,t2,t3,gamay,lambdax,w,Energy_midband,B)
syms B
RZ=t2*B^(-1)+t1+gamay+t3*B;
RF=t2*B+t1-gamay+t3*B^(-1);

parfor k=1:size(Energy_midband,2)
	FB=det([w-Energy_midband(1,k) RZ 0 0.5*lambdax 0 0; %det[H(beta)-E]
		RF w-Energy_midband(1,k) 0.5*lambdax 0 0 0;
		0 0.5*lambdax -Energy_midband(1,k) RZ 0 0.5*lambdax;
		0.5*lambdax 0 RF -Energy_midband(1,k) 0.5*lambdax 0;
		0 0 0 0.5*lambdax -w-Energy_midband(1,k) RZ;

		0 0 0.5*lambdax 0 RF -w-Energy_midband(1,k)]);
	Bb=solve(FB==0,B);
	E_beta_all(:,k)=double(Bb);
end
mm=0;
for k=1:size(E_beta_all,2)
	kkk=0;
	for kk=1:size(E_beta_all,1)
		kkk=kkk+1;
		E_beta(kkk,k)=E_beta_all(kk,k);
	end
	E_beta_abs=abs(E_beta(:,k));E_beta_sort=sort(E_beta_abs);
	index_Beta_M=fix(length(E_beta_sort)/2);
	for kkk=1:size(E_beta,1) %select beta by |Beta_M|=|Beta_M+1|
		if abs(E_beta_sort(index_Beta_M)-E_beta_sort(index_Beta_M+1))<0.05...
			&& abs(abs(E_beta(kkk,k))-E_beta_sort(index_Beta_M))<0.05
			mm=mm+1;
			Cbeta(1,mm)=E_beta(kkk,k);
		end
	end
end
%sort by the angle of Cbeta from -pi to pi
angle_Cbeta=zeros;
for kk=1:size(Cbeta,2)
	angle_Cbeta(kk,1)=angle(Cbeta(1,kk));
end
for ii=size(Cbeta,2)-1:-1:1
	for jj=1:ii
		if angle_Cbeta(jj,1)>angle_Cbeta(jj+1,1)
			temp_angle=angle_Cbeta(jj,1);
			angle_Cbeta(jj,1)=angle_Cbeta(jj+1,1);
			angle_Cbeta(jj+1,1)=temp_angle;
			temp_Cbeta=Cbeta(1,jj);
			Cbeta(1,jj)=Cbeta(1,jj+1);
			Cbeta(1,jj+1)=temp_Cbeta;
		end
	end
end
end
\end{lstlisting}

\end{document}